\documentclass[12pt]{article}
\pdfoutput=1
\usepackage{draft}
\usepackage{comment}
\usepackage[normalem]{ulem}
\usepackage{hyperref}
\hypersetup{
 colorlinks=true,
 urlcolor=blue,
 anchorcolor=blue,
 citecolor=blue,
 filecolor=blue,
 linkcolor=blue,
 menucolor=blue,
 linktocpage=true,
}
\usepackage{graphicx,color,subfig}
\usepackage{cite}
\usepackage{empheq} 
\usepackage[font=small,labelfont=bf]{caption} 
\usepackage{float}

\usepackage[dvipsnames,table]{xcolor}
\usepackage{tikz}
\usepackage{empheq}

\def\g{\gamma}

\usepackage{scalerel}
\usepackage{stackengine,wasysym}

\DeclareFontFamily{OT1}{pzc}{}
\DeclareFontShape{OT1}{pzc}{m}{it}{<-> s * [1.10] pzcmi7t}{}
\DeclareMathAlphabet{\mathpzc}{OT1}{pzc}{m}{it}

\definecolor{vert}{rgb}{0.1367 0.543 0.1367}
\def\De{\Delta}
\def\de{\delta}
\def\cO{\mathcal{O}}

\def\({\left(}
\def\){\right)}
\newcommand{\nn}{\nonumber}

\begin{document}

\unitlength = .8mm

\begin{titlepage}

\begin{center}

 \hfill \\
 \hfill \\

\title{Properties of scalar partition functions of 2d CFTs}

~\vskip 0.01 in

\author{
	Nathan Benjamin$^{a}$, Cyuan-Han Chang$^{b}$, A. Liam Fitzpatrick$^{c}$, Tobi Ramella$^{a}$
}
~\vskip 0.05 in

\emph{\small $^a$ Department of Physics and Astronomy \\
University of Southern California,
Los Angeles, CA 90089, USA \\ $^b$ Enrico Fermi Institute \& Kadanoff Center for Theoretical Physics \\
University of Chicago, Chicago, IL 60637, USA \\ $^c$ Department of Physics, Boston University, Boston, MA 02215, USA} 

~\vskip .2 in
\email{
	nathanbe@usc.edu, cchang10@uchicago.edu, fitzpatr@bu.edu, ramella@usc.edu 
}

\end{center}

\abstract{

We study the spectrum of scalar primary operators in any two-dimensional conformal field theory. We show that the scalars alone obey a nontrivial crossing equation. This extends previous work that derived a similar equation for Narain conformal field theories. Additionally, we show that at high temperature, the difference between the true scalar partition function and the one predicted from a semiclassical gravity calculation is controlled by: the modular integral of the partition function, the light states of the theory, and an infinite series terms directly related to the nontrivial zeros of the Riemann zeta function. We give several numerical examples and compute their modular integrals.

}

\vfill

\end{titlepage}

\eject

\begingroup

\baselineskip .168 in
\tableofcontents

\endgroup

\section{Introduction}
\label{sec:intro} 

Two-dimensional conformal field theories (CFTs) are of interest for a wide variety of reasons, including critical phenomena, worldsheet string theory, holography, and pure mathematics. Associated to every compact 2d CFT is its torus partition function, which is a modular invariant function:
\begin{equation}
    Z(\tau,\bar\tau) \coloneqq \sum_{j \in \mathbb Z} \int_{|j|-\frac{c}{12}}^\infty d \Delta \rho(\Delta, j) e^{-2\pi \Delta y} e^{2\pi i j x },
\end{equation}
where $\tau \coloneqq x + i y$, and $\rho(\Delta, j)$ is the density of states at energy $\Delta$ and spin $j$ (our conventions are that the vacuum has spin $j=0$ and energy $\Delta = -\frac{c}{12}$).

It is convenient to define the reduced partition function
\begin{align}
    Z_{p}(\tau,\bar\tau) &\coloneqq y^{1/2} |\eta(\tau)|^2 Z(\tau,\bar\tau) \nonumber \\ 
    &\coloneqq \sqrt y \sum_{j\in \mathbb Z}\int_{|j|-\frac{c-1}{12}}^\infty d\Delta \rho^p(\Delta, j) e^{-2\pi \Delta y} e^{2\pi i j x},
\end{align}
where $\rho^p(\Delta, j)$ (roughly) counts Virasoro primary operators of energy and spin $\Delta$ and $j$ respectively.\footnote{The vacuum in $\rho^p(\Delta,j)$ is now at $\Delta = -\frac{c-1}{12}$. Note that for states, we define $\Delta$ with a shift of $\frac{c}{12}$, and for primaries we define $\Delta$ with a shift of $\frac{c-1}{12}$. Due to the null state structure of the Virasoro vacuum character, $\rho^p(\Delta,j)$ is slightly different from the true density of Virasoro primary operators, and formally includes additional states with possibly negative degeneracy to take this into account. We will discuss this more in Sec. \ref{sec:Derivecross}.}

In this paper, we are especially interested in the \emph{scalar} primary operators, namely
\begin{align}
    \rho^{\text{scalars}}_p(\Delta) &\coloneqq \rho^p(\Delta, j=0),
\end{align}
which by definition obey\begin{equation}
    \int_{-\frac{c-1}{12}}^{\infty} d\Delta\, \rho^{\text{scalars}}_p(\Delta) e^{-2\pi \Delta y} = \int_{-1/2}^{1/2} dx\, Z(\tau,\bar\tau) |\eta(\tau)|^2.
\end{equation} 

Scalar primary operators are particularly interesting for a variety of reasons. First, the Lorentz-invariant renormalization group flows and conformal manifold starting from the CFT in the UV are determined, respectively, by the relevant and marginal scalar primary operators in the theory. In boundary CFT, scalars also play a privileged role in certain crossing equations as well. If one is interested in more refined information about black hole entropy for holographic theories, such as the entropy of AdS-Schwarzchild black holes, the scalars are the important states there too. 

Our main result in this paper is that the scalars of any 2d CFT obey very nontrivial crossing equations on their own. This generalizes previous work \cite{Benjamin:2022pnx} in which it was shown that the scalar $U(1)^c$ operators obey nontrivial crossing equations for any Narain CFT. Here, we generalize this to show that the \emph{Virasoro} scalar primary operators for \emph{any} 2d CFT obey nontrivial crossing equations.

We will show two separate crossing equations the scalar primaries obey. Our first result is the following. Define the function
\be
f(\Delta, r) \coloneqq \frac{\log(1-e^{-2\sqrt 2 \pi r \sqrt{\Delta}})}{r - r^{-1}}. 
\label{eq:fs1def}
\ee
We will show that for all $r$, any two-dimensional CFT obeys
\be
\int_{-\frac{c-1}{12}}^\infty d\Delta \rho^{\text{scalars}}_p(\Delta) \left[f(\Delta, r) + f(\Delta, r^{-1})\right] = \frac{\pi \varepsilon}6,
\label{eq:maincrossing}
\ee
where the RHS $\varepsilon$ is $r$-independent, and will be interpreted as a (regulated) integral of the partition function over the fundamental domain of $\mathbb{H}/SL(2,\mathbb Z)$. 

Taking derivatives of (\ref{eq:maincrossing}) with respect to $r$ gives a positive sum rule for Virasoro scalar primary operators, which allows us to bound the scalar gap of a generic 2d CFT. This then gives a bound on the scalar gap, which is more refined information than the usual (spinless) modular bootstrap equations \cite{Hellerman:2009bu, Friedan:2013cba, Benjamin:2016fhe, Collier:2016cls, Afkhami-Jeddi:2019zci, Hartman:2019pcd, Afkhami-Jeddi:2020hde}. Interestingly, unlike most bootstrap crossing equations, for sufficiently large central charge $c$, there does not exist a $r$ such that the LHS converges. More precisely, at large $\Delta$, the density of states $\rho^{\text{scalars}}_p(\Delta)$ grows as the Cardy formula \cite{Cardy:1986ie}
\begin{equation}
    \rho^{\text{scalars}}_p(\Delta) \sim \exp\left(2\pi\sqrt{\frac{(c-1)\Delta}{3}}\right).
\end{equation}
However, at large $\Delta$ we also have
\begin{equation}
    f(\Delta, r) + f(\Delta, r^{-1}) \sim \exp\left(-2\sqrt 2 \pi \sqrt\Delta ~\text{min}(r, r^{-1})\right)
\end{equation}
which means (\ref{eq:maincrossing}) converges for 
\begin{equation}
    c < 1+6(\text{min}(r,r^{-1}))^2 \leq 7.
\label{eq:needforconverge}
\end{equation}
For values of the central charge outside of (\ref{eq:needforconverge}), (\ref{eq:maincrossing}) must be defined via an analytic continuation in $r$ in the complex right-half-plane, namely by considering integrating $\rho^{\text{scalars}}_p(\Delta)$ against $f(\Delta, r)$ for $r$ with sufficiently large real part, and then analytically continuing to the whole complex right-half-plane, where it will then match integrating against $f(\Delta, r^{-1})$.

Our second main result is a more intricate crossing equation that scalar primaries obey. Roughly speaking, it takes the following form:
\begin{align}
    Z^{\text{scalars}}_p(y) - Z^{\text{gravity}}_{j=0}(y) &= \varepsilon + \sum_{k=1}^\infty \text{Re}\left(\delta_k y^{\frac{1+z_k}2}\right) \nonumber \\& + (\text{perturbative terms fixed by light spectrum})(y) + O(e^{-\#/y}),
\label{eq:schematicwow}
\end{align}
where: 
\begin{itemize}
    \item $Z^{\text{scalars}}_p(y) = \int_{-1/2}^{1/2} dx\, Z_p(\tau,\bar\tau)$, i.e. the scalar part of the reduced partition function.
    \item $Z^{\text{gravity}}_{j=0}(y)$ is the scalar part of the Poincar\'{e} sum of all light primary operators, those with $\Delta \leq 0$. Explicit expressions for light states of arbitrary spin can be found in \cite{Maloney:2007ud, Keller:2014xba}, and using this we will write a precise definition of $Z^{\text{gravity}}_{j=0}(y)$ in Sec. \ref{sec:Derivecross}. For holographic theories, this has an interpretation of a semiclassical gravity path integral; hence our name $Z^{\text{gravity}}$, but we emphasize that (\ref{eq:schematicwow}) is true for all 2d CFTs, not just holographic theories.
    \item $\varepsilon$ and $\delta_k$'s are theory-dependent constants (i.e. they do not depend on  $y$).
    \item $z_k$ are the nontrivial zeros of the Riemann zeta function with positive imaginary part. The Riemann hypothesis states that $\text{Re}(z_k) = \frac12$ for all $k$.
    \item There is an infinite series of perturbative error terms  that are exactly controlled by all light primaries. An explicit expression for these terms are given in (\ref{eq:crosswithzetastuff0}). The remaining error terms are non-perturbative in $y^{-1}$ and are controlled by all scalar primaries.
\end{itemize}
We give a precise, exact form of this equation in (\ref{eq:crosswithzetastuff0}), which includes the exact form of the error terms (both perturbative and non-perturbative). Unlike (\ref{eq:maincrossing}), (\ref{eq:schematicwow}) is convergent for all $y$ (with $\text{Re}(y)>0$). Applying a linear functional to (\ref{eq:schematicwow}) reproduces (\ref{eq:maincrossing}), so it is in some sense a more powerful crossing equation. However, due to the lack of positive-definiteness of the constants $\varepsilon, \delta_k$,  (\ref{eq:maincrossing}) is more convenient for bootstrap-type analyses.

A very natural question is: what is the asymptotic density of Virasoro scalar primary operators in a 2d CFT? The zeroth order answer comes from Cardy's formula. Additionally, there are an infinite set of corrections to Cardy's formula coming from other $SL(2,\mathbb Z)$ saddles, and other light states. Together, these form $Z^{\text{gravity}}(y)$. A remarkable feature of our result (\ref{eq:schematicwow}) is that it shows that after all of these corrections, the deviations come from a constant term ($\varepsilon$), and an infinite tower of oscillating terms controlled by the nontrivial zeros of the Riemann zeta functions. Their envelope has an overall size $y^{3/4}$ at small $y$, if the Riemann hypothesis is true.\footnote{The converse statement also holds assuming that for every value of $k$ there exists at least one CFT with $\delta_k \ne 0$.} 

The term $\varepsilon$ also has a nice physical interpretation. As we will soon see, it is the regulated modular integral of the entire partition function:
\begin{equation}
    \varepsilon = \frac3\pi \int_{\mathcal{F}} \frac{dx dy}{y^2} Z_p(\tau,\bar\tau),
\end{equation}
where $\mathcal{F}$ is the fundamental domain of $\mathbb H/SL(2,\mathbb Z)$. Thus, as a side benefit, we point out that (\ref{eq:schematicwow}) offers a computationally efficient method for computing a large class of integrals of modular invariants over the fundamental domain. These integrals show up often in the context of string theory; knowing the scalar sector of the theory lets us compute it efficiently. See e.g. \cite{Baccianti:2025gll} for recent work on a different computational method for computing integrals of this form, using a modular inversion formula. In fact there is a natural generalization of (\ref{eq:maincrossing}) for any power of $y$, which we discuss later.

This paper is organized as follows. In Sec.  \ref{sec:freebosonintegral}, we give a quick and intuitive derivation of one of our two main results, (\ref{eq:maincrossing}). In Sec. \ref{sec:Derivecross}, we give a derivation of our second main result, which is schematically (\ref{eq:schematicwow}). We also show how to derive (\ref{eq:maincrossing}) from (\ref{eq:schematicwow}). In Sec. \ref{sec:examples}, we give examples of several CFTs and show they obey (\ref{eq:maincrossing}). In Sec. \ref{sec:poles}, we discuss the pole structure of (\ref{eq:maincrossing}) in the complex $r$ plane. In Sec. \ref{sec:numericswithzetas}, we give numerical examples of (\ref{eq:schematicwow}). In Sec. \ref{sec:numerics}, we use bootstrap methods using semidefinite programming to get bounds on the scalar gap for $c<7$. Finally we discuss future directions in Sec. \ref{sec:discuss}.

\section{Fast Derivation of Crossing Equation}
\label{sec:freebosonintegral}

Consider two different `seed' functions of $y$ that resum, under their $SL(2,\mathbb{Z})/\Gamma_\infty$ images, into the same modular-invariant function $f(x,y)$:
\begin{equation}
    f(x,y) = \sum_{\gamma\in SL(2,\mathbb{Z})/\Gamma_\infty} \gamma\cdot  f_{\rm seed}(y) = \sum_{\gamma\in SL(2,\mathbb{Z})/\Gamma_\infty} \gamma\cdot \tilde{f}_{\rm seed}(y), \quad f_{\rm seed}(y) \ne \tilde{f}_{\rm seed}(y) .
    \label{eq:moneyfseed}
\end{equation}
Given a pair of such seed functions and a modular-invariant partition function $Z_p(x,y)$, one can take the Petersson inner product of $Z_p(x,y)$ and $f(x,y)$ and unfold it two different ways to get a nontrivial constraint on the scalar spectrum of $Z_p$:
\begin{equation}
    \int_0^\infty \frac{dy}{y^2} Z_{p}^{\text{scalars}}(y) (f_{\rm seed}(y) -\tilde{f}_{\rm seed}(y))= \int_{\mathcal{F} }\frac{dx dy}{y^2} Z_p(x,y) (f(x,y)-f(x,y))  = 0,
    \label{eq:GeneralScalarCrossingEq}
    \end{equation}
    provided that the integral converges.

    One example of such a pair of seed functions arises from the fact that the torus partition function of the $c=1$ free compact boson with radius $r$ is the modular image of the Jacobi Theta function
    \begin{equation}
        \hat{Z}^{(c=1)}(x,y;r) = r \sum_{\gamma \in SL(2,\mathbb{Z})/\Gamma_\infty} \gamma \cdot \vartheta(r^2/y), \qquad \vartheta(t) \equiv \sum_{m=-\infty}^\infty e^{-\pi m^2 t } = \frac{1}{\sqrt{t}} \vartheta(\frac{1}{t}) .
\end{equation}
This implies that the Theta function is a seed function for the compact free boson partition function, and its duality under $r\rightarrow 1/r$ relates this seed to another seed function $\tilde{f}$:
\begin{equation}
    f_{\rm seed}(y) =  r \vartheta(r^2/y), \qquad \tilde{f}_{\rm seed}(y) = r^{-1} \vartheta(r^{-2}/y).
\end{equation}
However, the constant term $m=0$ in both sums  causes the integral in (\ref{eq:GeneralScalarCrossingEq}) to diverge.  To regulate it, we can add and subtract the $m=0$ term in the sum in the defining equation for $\vartheta$:
\begin{equation}
    f_{\rm seed}(y) - \tilde{f}_{\rm seed}(y) \rightarrow f_{\rm seed}(y) - \tilde{f}_{\rm seed}(y) - (r-r^{-1}).
\end{equation}
If we modify the integrand (\ref{eq:GeneralScalarCrossingEq}) this way, then the RHS will no longer be zero, but instead will be an unknown quantity given by integrating $Z_{p}^{\text{scalars}}(y)$ against the additional terms.  At this point we can take advantage of the fact that the $c=1$ compact boson provides a family of seed function pairs as a function of $r$, and as long as the unknown function is independent of $r$, we can just differentiate the equation with respect to $r$ to get a useful constraint on the scalar spectrum of $Z_p$. To render the unknown function independent of $r$, we divide the integrand by the $r$ dependence of the regulating term.  That is, we define
\begin{equation}
\begin{aligned}
    \frac{\pi}{3} \varepsilon &\equiv  -\int_0^\infty \frac{dy}{y^2} Z_{p}^{\text{scalars}}(y) \Big(\frac{f_{\rm seed}(y;r) -\tilde{f}_{\rm seed}(y;r)}{r-r^{-1}} - 1\Big) .
    \label{eq:EpsDef}
    \end{aligned}
\end{equation}
This integral is well-defined for $c<1+6\textrm{min}(r^2,\frac{1}{r^2})$, since as we will see shortly it converges term-by-term in the sum over scalar states in $Z_{p}$, and the sum over scalar states converges for $c<1+6\textrm{min}(r^2,\frac{1}{r^2})$ due to the asymptotic Cardy density of scalar states.  A non-rigorous way to argue that it is independent of $r$ is that when one `folds' the integral back up into an integral over the fundamental domain $\mathcal{F}$, the two seed terms both sum up to the same function and cancel, whereas the regulating term ``$-1$'' is independent of $r$.  However, separating the folded integral over $\mathcal{F}$ into these two distinct pieces is more involved, because the integral over the regulating term alone is divergent.  Therefore, when we do the folding, we must regulate the integrals in some way, and then it is necessary to show that the regulator does not reintroduce additional dependence on $r$. We will describe how to do this below, but for the moment let us simply assume that it is possible to do the necessary regularization without introducing additional $r$ dependence.\footnote{We emphasize that expression (\ref{eq:EpsDef}) is already a convergent definition for $\varepsilon$ as long as $c<1+6 \textrm{min}(r^2,1/r^2)$, so for any given $Z_{p}^{\text{scalars}}(y)$, whether  $\varepsilon$ depends on $r$ or not is not dependent on choices of regulators.  The issue is whether or not we can find an expression for the RHS that is manifestly independent of $r$. }

Next, we can process this constraint into a more usable form by doing the integral over $y$ term-by-term in the sum over scalar states in $Z_p$:
\begin{equation}
\int_0^\infty \frac{dy}{y^{3/2}} e^{-2 \pi y \Delta} r (\vartheta(r^2/y)-1)= 2\sum_{m=1}^\infty \frac{e^{-2\pi\sqrt{2} m  r \sqrt{\Delta}}}{m} = - 2 \log(1-e^{-2 \sqrt{2} \pi r \sqrt{\Delta}}).
\end{equation}
Collecting terms and simplifying, we find
\begin{equation}
   \frac{\pi}{6} \varepsilon = -\int_{-\frac{c-1}{12}} ^\infty d\Delta \rho^{\rm scalars}(\Delta) \frac{  \log(1-e^{-2 \sqrt{2} \pi r \sqrt{\Delta}}) - \log(1-e^{-2 \sqrt{2} \pi r^{-1}  \sqrt{\Delta}})}{r^{-1} -r} ,
\end{equation}
 which is exactly the scalar crossing equation (\ref{eq:maincrossing}). 

Finally, we return to the issue of regulating the integral over the ``$-1$'' piece in (\ref{eq:EpsDef}).  Roughly, we would like to fold the integral over $y$ back into an integral over the fundamental domain by summing over $SL(2,\mathbb{Z})$ images of the integrand, to get an integral of a modular invariant function over the fundamental domain, as follows:
\begin{equation}
   \int_0^\infty \frac{dy}{y^2}  Z_{p}^{\text{scalars}}(y) \Big(\frac{f_{\rm seed}(y;r) -\tilde{f}_{\rm seed}(y;r)}{r-r^{-1}} -1\Big) 
    \stackrel{?}{=} \int_{\mathcal{F}}\frac{dx dy}{y^2}\Big( Z_p \frac{Z^{(c=1)}(r)- Z^{(c=1)}(r^{-1})}{r-r^{-1}} - Z_p \Big),
     \label{eq:RegRS}
\end{equation}
so that we can cancel the term $Z^{(c=1)}(r)-Z^{(c=1)}(r^{-1})$.  This is obviously not allowed, though, because the integral of $Z_p$ over the fundamental domain diverges exponentially at $y \rightarrow \infty$ due to the light states, $\Delta < 0$, so the RHS of the expression above is infinite whereas the LHS is finite. What we can do instead is define a subtracted partition function 
\begin{equation}
    \hat{Z}_p = Z_p - \hat{Z}_L,
\end{equation}
where $\hat{Z}_L$ is a modular-invariant partition function of our choosing with the same light spectrum as $Z_p$. Then, separating $Z_p = \hat{Z}_p + \hat{Z}_L$  in the folded-up integral over $\mathcal{F}$, we instead obtain
\begin{equation}
    \frac{\pi}{3} \varepsilon = \int_{\mathcal{F}} \frac{dx dy}{y^2} \hat{Z}_p - \int_0^\infty \frac{dy}{y^2}  \hat{Z}_{L,0}(y) \Big(\frac{f_{\rm seed}(y;r) -\tilde{f}_{\rm seed}(y;r)}{r-r^{-1}} -1\Big) .
\end{equation}
The first term on the RHS is now finite and manifestly independent of $r$.  In Section \ref{sec:polestructuremwk} and Appendix \ref{sec:putcalchere}, we show by explicit calculation that if we choose $\hat{Z}_L$ to be the MWK completion of the light scalar states in $Z_p$, then the second term is also independent of $r$ (and moreover only gets contributions from the completion of the light scalar states in $Z_p$). Therefore, $\varepsilon$ is independent of $r$, as we wished to show.

It would be interesting if there were other choices of functions $f_{\text{seed}}(y), \tilde f_{\text{seed}}(y)$ that obeyed a similar equation as (\ref{eq:moneyfseed}). In particular it would be nice if there were a choice that led to a faster falloff, and allowed for a nontrivial crossing equation at all $c$. A natural guess is integrating against a Narain theta function at general central charge instead of just at $c=1$, but unfortunately at higher $c$, the full theta function cannot be written as a simple Poincare sum of scalar states (though at $c=2$ there is an interesting interpretation in terms of single-trace $T\bar{T}$-deformations, see e.g. \cite{Callebaut:2019omt}).

\section{Second Derivation of Crossing Equation}
\label{sec:Derivecross}

In this section, we give a derivation for the two scalar crossing equations, \eqref{eq:maincrossing} and \eqref{eq:schematicwow}. We start with a very brief review of harmonic analysis in section \ref{sec:harmonic}, which will be our main tool for deriving the crossing equation. We then explain the derivation of \eqref{eq:schematicwow} in section \ref{sec:crossing_derivation}. In section \ref{sec:functional_sumrules}, we apply linear functionals to the equation to obtain \eqref{eq:maincrossing}, which will be used later to obtain bounds on the scalar gap.

\subsection{Review of harmonic analysis}\label{sec:harmonic}

The discussion in this subsection closely follows \cite{Terras_2013,Benjamin:2021ygh}. Every modular-invariant function $f(\tau,\bar\tau)$ that is square integrable, i.e.
\begin{equation}
    \int_{\mathcal{F}} \frac{dx dy}{y^2} f(\tau,\bar\tau) < \infty,
\end{equation}
admits a unique decomposition into eigenfunctions of the Laplacian. This is of the form
\begin{equation}
    f(\tau,\bar\tau) = \frac 3\pi (f, 1) + \frac{1}{4\pi i} \int_{\text{Re}(s) = \frac12} (f, E_s) E_s(\tau,\bar\tau) + \sum_{i=1}^\infty \sum_{\pm} \frac{\nu_i^\pm(\tau,\bar\tau) (f, \nu_i^\pm)}{(\nu_i^\pm, \nu_i^\pm)},
\label{eq:specdecomp}
\end{equation}
where 
\begin{equation}
(f, g) \coloneqq \int_{\mathcal{F}} \frac{dx dy}{y^2} f(\tau,\bar\tau) g(\tau,\bar\tau)^*,
\label{eq:defofoverlap}
\end{equation}
and $E_s(\tau,\bar\tau)$ are real analytic Eisenstein series, and $\nu_i^\pm(\tau,\bar\tau)$ are Maass cusp forms; both of them are eigenfunctions of the Laplacian. In this paper, we will focus on the scalar part of these functions, which are:\footnote{In this work, by scalar part of a function $f(\tau,\bar \tau)$ we mean its zeroth Fourier mode,
\begin{equation}
    \int_{-1/2}^{1/2}dx\, f(\tau,\bar \tau),
\end{equation}
where $x=\mathrm{Re}(\tau)$. For partition functions this projects onto the contributions of the scalar operators. By contrast, the other Fourier components will be referred to as the spinning part.
}
\begin{align}
E_s(\tau,\bar\tau) &= y^s + \frac{\Lambda(1-s)}{\Lambda(s)}y^{1-s} + \sum_{j\neq 0} e^{2\pi i j x} (\ldots) \nonumber \\
\nu^\pm_i(\tau,\bar\tau) &= \sum_{j\neq 0} e^{2\pi i j x} (\ldots),
\end{align}
with the function $\Lambda(s) \coloneqq \pi^{-s} \zeta(2s)\Gamma(s)$.
(See e.g. \cite{Terras_2013,Benjamin:2021ygh} for full definitions of the Eisenstein series and Maass  cusp forms.) Note that the Eisenstein series satisfies the following functional equation
\begin{equation}\label{eq:Es_identity}
    \Lambda(s)E_s(\tau,\bar \tau) = \Lambda(1-s)E_{1-s}(\tau,\bar \tau).
\end{equation}

More generally, in \cite{zagier1981rankin} Zagier showed that this method can be extended to functions of polynomial growth at $y\to \infty$. Consider a function $f(\tau,\bar \tau)$ that behaves like
\begin{equation}
    f(\tau,\bar \tau) = \sum_{i}c_i y^{\alpha_i} + (\text{sub-polynomial}),\quad y\to \infty.
\end{equation}
Then, we can define
\begin{equation}
    \tilde f(\tau,\bar \tau) \equiv f(\tau,\bar \tau) - \sum_{i|\alpha_i \geq 1/2} c_i E_{\alpha_i}(\tau,\bar \tau),
\end{equation}
which is now a square-integrable function. The spectral decomposition of $f(\tau,\bar \tau)$ is then given by
\begin{equation}
    f(\tau,\bar \tau) = \sum_{i|\alpha_i \geq 1/2} c_i E_{\alpha_i}(\tau,\bar \tau) + \frac 3\pi (\tilde f, 1) + \frac{1}{4\pi i} \int_{\text{Re}(s) = \frac12} (\tilde f, E_s) E_s(\tau,\bar\tau) + \sum_{i=1}^\infty \sum_{\pm} \frac{\nu_i^\pm(\tau,\bar\tau) (\tilde f, \nu_i^\pm)}{(\nu_i^\pm, \nu_i^\pm)}.
    \label{eq:specdecomp_2}
\end{equation}

If we take the scalar part of (\ref{eq:specdecomp_2}), i.e. integrate in $x$ over a strip, then by definition we project onto the scalar parts of each of the eigenfunctions. The Maass cusp forms have no scalar part, and the real analytic Eisenstein series have a very simple scalar part. After using the $s \leftrightarrow 1-s$ symmetry and \eqref{eq:Es_identity}, we get:
\begin{equation}
\int_{-1/2}^{1/2} dx f(\tau,\bar\tau) = \sum_{i|\alpha_i \geq 1/2} c_i \left(y^{\alpha_i}+\frac{\Lambda(\alpha_i)}{\Lambda(1-\alpha_i)}y^{1-\alpha_i}\right)+\frac3\pi (\tilde f,1) + \frac1{2\pi i}\int_{\text{Re(s)}=\frac12} (\tilde f,E_s) y^s.
\label{eq:fscalar}
\end{equation}
We would like to apply (\ref{eq:fscalar}) for a 2d CFT partition function. To do so, we first must subtract out all light states, which make the partition function grow exponentially at $y\to \infty$, in a modular invariant way.

\subsection{Derivation of the crossing equation}\label{sec:crossing_derivation}
\subsubsection{Subtracting the light spectrum}

For a generic 2d CFT with $c>1$, we consider its primary-counting partition function $Z_p$ defined as
\be
Z_p(\tau,\bar \tau) = y^{\frac{1}{2}}|\eta(\tau)|^2 Z(\tau, \bar \tau) = y^{\frac{1}{2}}\left(q^{-\xi}\bar q^{-\xi}|1-q|^2+\sum_{\mathrm{primary}}q^{h-\xi}\bar q^{\bar h-\xi}\right),
\ee
where $\xi=\frac{c-1}{24}$, $q=e^{2\pi i \tau}$, and $\eta(\tau)$ is the Dedekind eta function. Unlike the usual partition function $Z(\tau,\bar \tau)$ which counts both primaries and descendants, $Z_p(\tau,\bar\tau)$ only contains primary operators in its sum. We have also written the contribution from vacuum separately due to the null states.

Even though $Z_p$ is a modular-invariant function, it is not square-integrable. Indeed, at the cusp $y\to \infty$ we find
\be
Z_p \sim y^{\frac{1}{2}}e^{4\pi y\xi},\qquad y\to \infty,
\ee
and the exponential growth makes the integrals (\ref{eq:defofoverlap}) divergent. The leading behavior at the cusp is due to the vacuum, but in fact any operator with $\De=h+\bar h-2\xi<0$ will lead to an exponential growth in the $y\to \infty$ limit. For later convenience, let us define
\begin{align}
\cL &=\left\{\textrm{scalar primaries with } \Delta \leq 0 \right\},\qquad \mathcal{S} =\left\{\textrm{scalar primaries with } \Delta > 0\right\},\nn \\
\mathcal{J} &=\left\{\textrm{spinning primaries with } \Delta \leq 0 \right\}.
\end{align}
We will also use $\cL'$ to denote the set $\cL$ with vacuum excluded. As long as $\cL \cup \mathcal{J}$ is nonempty, $Z_p$ will diverge exponentially at $y \to \infty$.

To obtain a square-integrable function, the authors of \cite{Benjamin:2021ygh} propose that one can try to remove the contributions of the light operators $\cL \cup \mathcal{J}$ from $Z_p$ in a modular-invariant way. In other words, one should look for a modular-invariant function $\hat Z_L$ that has the same light spectrum as $Z_p$, and consider a subtracted partition function $\hat Z_p$,
\be
\hat{Z}_p(\tau,\bar \tau) = Z_p(\tau,\bar \tau) - \hat Z_L(\tau,\bar \tau).
\ee
Note that the choice of $\hat Z_L$ is not unique. Even if the light spectrum of $\hat Z_L$ is fixed, there are different ways of completing it to make it a modular-invariant function. In this paper, we consider a choice that seems the most natural and over which we also have good analytic control. It is built out of the Maloney-Witten-Keller (MWK) partition function \cite{Maloney:2007ud, Keller:2014xba}, given by replacing each light operator $\cO\in \cL \cup \mathcal{J}$ with its Poincar\'{e} sum. For a light operator $\cO$ with energy $E$ and spin $J$, the MWK partition function is given by
\begin{align}
Z^{\mathrm{MWK}}_{E,J}(\tau,\bar \tau) \equiv& \sum_{\g \in SL(2,\mathbb{Z})/\Gamma_{\infty}} \left.\left(y^{\frac{1}{2}}q^{h-\xi}\bar q^{\bar h-\xi}\right)\right|_{\g} \nn \\
=&  \sum_{\g \in SL(2,\mathbb{Z})/\Gamma_{\infty}} \left.\left(y^{\frac{1}{2}} e^{-2\pi y E} e^{2\pi i x J}\right)\right|_{\g}.
\end{align}
When $\cO$ is the vacuum, we need a slightly different definition due to the null states,
\begin{align}\label{eq:MWKvac_decomposed}
Z^{\mathrm{MWK}}_{\mathrm{vac}}(\tau,\bar \tau) \equiv& \sum_{\g \in SL(2,\mathbb{Z})/\Gamma_{\infty}} \left.\left(y^{\frac{1}{2}}q^{-\xi}\bar q^{-\xi}(1-q-\bar q+q\bar q)\right)\right|_{\g} \nonumber \\
=&Z^{\mathrm{MWK}}_{-2\xi,0} - Z^{\mathrm{MWK}}_{-2\xi+1,1} - Z^{\mathrm{MWK}}_{-2\xi+1,-1} + Z^{\mathrm{MWK}}_{-2\xi+2,0}.
\end{align}
We record the expressions of the scalar part of $Z^{\mathrm{MWK}}_{E,J}$ in Appendix \ref{app:MWK_details}.

Using the MWK partition function, we can define our subtracted partition function $\hat Z_p$ to be
\begin{align}\label{eq:Zphat_definition}
\hat{Z}_p \equiv Z_p-Z^{\text{gravity}},
\end{align}
where
\begin{equation}
    Z^{\text{gravity}} = Z^{\mathrm{MWK}}_{\mathrm{vac}} + \sum_{\cO \in \mathcal{L}' \cup \mathcal{J}} Z^{\mathrm{MWK}}_{E,J}.
\end{equation}

From the expressions given in \eqref{eq:ZMWK_scalar} and \eqref{eq:ZMWK_spinning}, we see that at $y\to \infty$, the leading behavior of $\hat Z_p$ is\footnote{The spinning part of $Z^{\mathrm{MWK}}$ is exponentially small at large $y$.}
\be\label{eq:Zp_largey}
\hat Z_p \sim y^{\frac{1}{2}}\(6+\sum_{\cO\in \mathcal{L}'}1-\sum_{\cO\in\mathcal{J}}2\sigma_0(J)\) + O(y^{-\frac{1}{2}}),\qquad y\to \infty,
\ee
where $\sigma$ is the divisor sigma function, and the $6$ is from the four terms in vacuum contribution \eqref{eq:MWKvac_decomposed}. So, $\hat Z_p$ grows polynomially at the cusp $y\to \infty$, and as we discussed in the previous subsection, this is good enough for us to write down its spectral decomposition.

\subsubsection{Scalar crossing equation}
By \eqref{eq:specdecomp_2}, the spectral decomposition of $\hat Z_p$ is given by
\begin{align}\label{eq:Zphat_decomposition}
\hat Z_p(\tau) =& \frac{3}{\pi} (\hat Z_p,1)
+\frac{1}{4\pi i}\int_{\mathrm{Re}(s)=\frac{1}{2}}ds (\hat Z_p,E_s)E_s(\tau) + \mathrm{cusps},
\end{align}
where ``cusps" are contributions from the Maass cups forms. Note that polynomially growing term $y^{1/2}$ in \eqref{eq:Zp_largey} does not contribute since $E_{1/2}=0$. As we will see later, the effect of the $y^{1/2}$ term will appear in the pole of $(\hat Z_p,E_s)$.

Taking the scalar part of \eqref{eq:Zphat_decomposition}, the contributions from the cusp forms vanish, and we obtain
\begin{align}\label{eq:Vircross_expr0}
&Z^{\text{scalars}}_p-Z^{\text{gravity}}_{j=0}(y)
=\frac{3}{\pi} (\hat Z_p,1)+\frac{1}{2\pi i}\int_{\mathrm{Re}(s)=\frac{1}{2}}ds\, (\hat Z_p,E_s)y^s,
\end{align}
where $Z^{\text{gravity}}_{j=0}(y)=\int_{-1/2}^{1/2}dx\, Z^{\text{gravity}}(\tau,\bar \tau)$ is the scalar sector of $Z^{\text{gravity}}$, and $Z^{\text{scalars}}_p$ is the scalar partition function,
\begin{equation}
    Z^{\text{scalars}}_p =y^{\frac{1}{2}}\(e^{4\pi y\xi}+e^{4\pi y(\xi-1)}+\sum_{\cO \in \mathcal{L}' \cup \mathcal{S}}e^{-2\pi y \De}\).
\end{equation}

The left-hand side of \eqref{eq:Vircross_expr0} depends on the scalar operators and the light spinning operators of the CFT. On the other hand, on the right-hand side the theory-dependent terms are $(\hat Z_p,1)$ and $(\hat Z_p,E_s)$, the overlap of $\hat{Z}_p$ with the Eisenstein series. Our next step is to compute the $s$-integral on the right-hand side. To do this, first note that due to the functional equation \eqref{eq:Es_identity}, we have
\begin{equation}\label{eq:ZpEs_identity}
    (\hat Z_p,E_s) = \frac{\Lambda(s)}{\Lambda(1-s)}(\hat Z_p,E_{1-s}) = \frac{\Gamma(s)\zeta(2s)}{\pi^{\frac{1}{2}}\Gamma(s-\frac{1}{2})\zeta(2s-1)}(\hat Z_p,E_{1-s}).
\end{equation}
Furthermore, $(\hat Z_p,E_{1-s})$ can be written as
\begin{align}\label{eq:ZpEs_overlap_eq0}
    (\hat Z_p,E_{1-s}) =& \int_{\mathcal{F}} \frac{dx dy}{y^2}\hat Z_p(\tau,\bar \tau) E_{s}(\tau,\bar \tau) \nn \\
    =&\int dy \int d\Delta \, \bigg(\rho^{\text{scalars}}_p(\Delta)-\sum_{\cO\in \widetilde{\mathcal{L}}}\rho^{\text{scalars}}_{\text{MWK},E,0}(\Delta)-\sum_{\cO\in \widetilde{\mathcal{J}}}\rho^{\text{scalars}}_{\text{MWK},E,J}(\Delta)\bigg)e^{-2\pi \Delta y}y^{s-\frac{3}{2}}.
\end{align}
From the first line to the second line we have plugged in \eqref{eq:Zphat_definition} and used the unfolding trick, where one writes $E_s$ as a Poincar\'{e} sum of $y^s$. The sets $\widetilde{\mathcal{L}}$ and $\widetilde{\mathcal{J}}$ are formally defined as
\begin{align}
    \widetilde{\mathcal{L}}&= \mathcal{L}' \cup \{\cO_{\De=-2\xi,J=0}, \cO_{\De=-2\xi+2,J=0}\}, \nn \\
    \widetilde{\mathcal{J}}&= \mathcal{J} \cup \{-\cO_{\De=-2\xi+1,J=1}, -\cO_{\De=-2\xi+1,J=-1}\},
\end{align}
where the negative sign means we should formally treat the degeneracy as $-1$.\footnote{In this notation, we have $\rho^{\text{scalars}}_p(\De) = \rho_{\widetilde{\mathcal{L}}\cup\mathcal{S}}(\De).$}

If we are allowed to exchange the two integrals in \eqref{eq:ZpEs_overlap_eq0}, then we can evaluate the $y$-integral and write $(\hat{Z}_p,E_{1-s})$ as a single integral over $\Delta$. In this paper, we will assume that we can do this for sufficiently large $\mathrm{Re}(s)$. In particular, we assume
\begin{equation}\label{eq:ZpEs_overlap_assumption}
    \rho^{\text{scalars}}_p(\Delta)-\sum_{\cO\in \widetilde{\mathcal{L}}}\rho^{\text{scalars}}_{\text{MWK},E,0}(\Delta)-\sum_{\cO\in \widetilde{\mathcal{J}}}\rho^{\text{scalars}}_{\text{MWK},E,J}(\Delta) = O(\Delta^{-1/2}),\quad \Delta \to \infty,
\end{equation}
which will allow us to exchange the integrals in \eqref{eq:ZpEs_overlap_eq0} for $\text{Re}(s)>1$. The intuition for this assumption comes from the fact that the $\rho^{\text{scalars}}_{\textrm{MWK}}$ terms subtract all the Cardy growth in $\rho^{\text{scalars}}_p(\Delta)$, and the remaining pieces should be subleading. In Sec. \ref{sec:numericswithzetas}, we provide numerical evidence for the assumption \eqref{eq:ZpEs_overlap_assumption} by checking it for various solvable CFTs, which are all consistent with the $O(\Delta^{-1/2})$ behavior. Note that although we checked this numerically for RCFTs in Sec. \ref{sec:numericswithzetas}, we expect the falloff in (\ref{eq:ZpEs_overlap_assumption}) to be even faster for chaotic CFTs, due to a lack of large degeneracies in $\rho^{\text{scalars}}_p(\Delta)$.\footnote{Less precisely, chaotic theories in some sense should have a density of scalars that is ``closer" to $\rho^{\text{scalars}}_{\text{MWK}}$ (which is a smooth function), since the spacing between energies should be exponentially small in energy, as opposed to quantized. For instance, in Sec. \ref{sec:numericswithzetas} we check the $(E_8)_1$ WZW theory, which has primaries at integer-spacing; this has been proven to be the sparsest possible \cite{Mukhametzhanov:2019pzy, Mukhametzhanov:2020swe}.} Note also that the assumption is in fact stronger than necessary, and the derivation would still hold as long as the LHS of \eqref{eq:ZpEs_overlap_assumption} does not grow superpolynomially.\footnote{For example, if we have $\rho^{\text{scalars}}_p(\Delta)-\sum_{\cO\in \widetilde{\mathcal{L}}}\rho^{\text{scalars}}_{\text{MWK},E,0}(\Delta)-\sum_{\cO\in \widetilde{\mathcal{J}}}\rho^{\text{scalars}}_{\text{MWK},E,J}(\Delta) = O(\Delta^{s_0-1/2})$, we can move the $s$-contour to $\text{Re}(s)=s_0+1$, and then plug in \eqref{eq:ZpEs_overlap_assumption_afteridentity}. Depending on the value of $s_0$, when moving the contour we might have to separate some terms in $\rho^{\text{scalars}}_{\text{MWK}}$ as explained in footnote \ref{footnoteonsmallDelta}, but the final result \eqref{eq:crosswithzetastuff0} is the same.}

Combining \eqref{eq:ZpEs_overlap_eq0}, the assumption \eqref{eq:ZpEs_overlap_assumption}, and the functional identity \eqref{eq:ZpEs_identity}, we have\footnote{From the assumption, the integral converges at large $\De$ for $\text{Re}(s)>1$. However, due to the $\rho^{\text{scalar}}_{\text{MWK}}$ terms in the integrand, the integral at small $\De$ converges only for $\text{Re}(s)<\frac{3}{2}$, and therefore one should only use this equation for $\frac{3}{2}>\text{Re}(s)>1$. For $\text{Re}(s)>\frac{3}{2}$, the divergent terms in $\rho^{\text{scalar}}_{\text{MWK}}$ should be treated separately. See appendix \ref{app:ZpEs_poles} for more discussions. \label{footnoteonsmallDelta}}
\begin{align}\label{eq:ZpEs_overlap_assumption_afteridentity}
    &(\hat{Z}_p,E_s) = \frac{2^{\frac{1}{2}-s}\Gamma(s)\zeta(2s)}{\pi^{s}\zeta(2s-1)} \nn \\
    &\qquad\qquad \times \int d\Delta\, \bigg(\rho^{\text{scalars}}_p(\Delta)-\sum_{\cO\in \widetilde{\mathcal{L}}}\rho^{\text{scalars}}_{\text{MWK},E,0}(\Delta)-\sum_{\cO\in \widetilde{\mathcal{J}}}\rho^{\text{scalars}}_{\text{MWK},E,J}(\Delta)\bigg)\Delta^{\frac{1}{2}-s},\quad \mathrm{Re}(s)>1.
\end{align}

Next, we can move the $s$-integral contour in \eqref{eq:Vircross_expr0} to the right to $\mathrm{Re}(s)=\gamma$ where $\gamma>1$. We will additionally choose $\gamma<\frac{3}{2}$ to  avoid divergences at small $\De$ (see footnote \ref{footnoteonsmallDelta}). When we deform the contour, we need to include contributions of the poles of $(\hat Z_p,E_s)$ between $\mathrm{Re}(s)=\frac{1}{2}$ and $\mathrm{Re}(s)=\gamma$. As we explain in appendix \ref{app:ZpEs_poles}, it turns out that the only poles are at $s=\frac{1}{2},\frac{1+z_k}{2}, \frac{1+z_k^*}{2}$, where $z_k$'s are the nontrivial zeros of the Riemann zeta function.\footnote{Near $s=\frac{1}{2}$, we slightly deform the contour such that the $s=\frac{1}{2}$ pole lies on the right of the contour. See appendix \ref{app:ZpEs_poles} for details.} (A quick way of seeing the poles at $s=\frac{1+z_k}{2}, \frac{1+z_k^*}{2}$ is by considering \eqref{eq:ZpEs_identity}, where they correspond to the zeros of $\zeta(2s-1)$ in the denominator.) Consequently, after deforming the contour and plugging in \eqref{eq:ZpEs_overlap_assumption_afteridentity}, we obtain
\begin{align}\label{eq:Vircross_expr1}
    &Z^{\text{scalars}}_p-Z^{\text{gravity}}_{j=0}(y) \nn \\
=&\frac{3}{\pi} (\hat Z_p,1)+ \sum_k \mathrm{Re}(\delta_k y^{\frac{1+z_k}{2}})-\(\sum_{\widetilde{\mathcal{L}}} 1 - \sum_{\widetilde{\mathcal{J}}} 2\sigma_0(J)\)y^{\frac{1}{2}} \nn \\
    +&\frac{1}{2\pi i}\int_{\mathrm{Re}(s)=\gamma}ds\, \frac{2^{\frac{1}{2}-s}\Gamma(s)\zeta(2s)}{\pi^{s}\zeta(2s-1)}\int d\Delta\, \bigg(\rho^{\text{scalars}}_p(\Delta)-\sum_{\cO\in \widetilde{\mathcal{L}}}\rho^{\text{scalars}}_{\text{MWK},E,0}(\Delta)-\sum_{\cO\in \widetilde{\mathcal{J}}}\rho^{\text{scalars}}_{\text{MWK},E,J}(\Delta)\bigg)\Delta^{\frac{1}{2}-s}y^s,
\end{align}
where $\delta_k=-\mathrm{Res}_{s=\frac{1+z_k}{2}}(\hat{Z}_p,E_s)$, and $\(\sum_{\widetilde{\mathcal{L}}} 1 - \sum_{\widetilde{\mathcal{J}}} 2\sigma_0(J)\)$ is the residue of the pole at $s=\frac{1}{2}$ (see \eqref{eq:overlap_polefromMWK}).

To compute the integral in the last line of \eqref{eq:Vircross_expr1}, we can consider the expansion
\begin{equation}
    \frac{\zeta(2s)}{\zeta(2s-1)} = \sum_{n=1}^\infty b(n)n^{-2s},
    \label{eq:bndef}
\end{equation}
where the coefficient $b(n)$ is given by $b(n)=\sum_{k|n} k\mu(k)$, where $\mu(k)$ is the M\"{o}bius function. We also have the following identity:
\begin{equation}
    \frac{1}{2\pi i}\int_{\mathrm{Re}(s)=\g} ds \frac{\Gamma(s)}{\pi^{s}}(2\De)^{\frac{1}{2}-s}y^{s}n^{-2s} = \sqrt{2\De} e^{-\frac{2\pi n^2\De}{y}}.
\end{equation}
Then, we have
\begin{align}\label{eq:Vircross_expr2}
    &Z^{\text{scalars}}_p(y)-Z^{\text{gravity}}_{j=0}(y) \nn \\
=&\frac{3}{\pi} (\hat Z_p,1)+ \sum_k \mathrm{Re}(\delta_k y^{\frac{1+z_k}{2}})-\(\sum_{\widetilde{\mathcal{L}}} 1 - \sum_{\widetilde{\mathcal{J}}} 2\sigma_0(J)\)y^{\frac{1}{2}} \nn \\
    +&\int d\Delta\, \bigg(\rho^{\text{scalars}}_p(\Delta)-\sum_{\cO\in \widetilde{\mathcal{L}}}\rho^{\text{scalars}}_{\text{MWK},E,0}(\Delta)-\sum_{\cO\in \widetilde{\mathcal{J}}}\rho^{\text{scalars}}_{\text{MWK},E,J}(\Delta)\bigg)\sum_{n=1}^{\infty}b(n) \sqrt{2\De} e^{-\frac{2\pi n^2\De}{y}}.
\end{align}

By plugging in \eqref{eq:rhoMWK_scalarseed} and \eqref{eq:rhoMWK_spinningseed}, we can compute the $\Delta$-integrals involving $\rho^{\text{scalars}}_{\text{MWK}}$ exactly. The final result is given by
\begin{align}\label{eq:crosswithzetastuff0}
    &Z_{p}^{\text{scalars}}(y) -Z^{\text{gravity}}_{j=0}(y) \nn \\
=&\frac{3}{\pi} (\hat Z_p,1)+ \sum_k \mathrm{Re}(\delta_k y^{\frac{1+z_k}{2}})  \nn \\
-&y^{\frac{1}{2}}\sum_{\cO\in\widetilde{\mathcal{L}}}e^{-2\pi \Delta y} -y^{\frac{1}{2}}\sum_{\cO\in\widetilde{\mathcal{J}}}\left(-2\sigma_{0}(|J|)+\sum_{m=1}^{\infty}\frac{2\pi^{m}\sigma_{2m}(J)T_m(\tfrac{-\Delta}{|J|})}{\Gamma(m+1)|J|^m\zeta(2m)}y^{m}\right) \nn \\
+&\sum_{\cO\in \mathcal{S}}\sum_{n=1}^{\infty} b(n) \sqrt{2\De} e^{-\frac{2\pi n^2\De}{y}}.
\end{align}
Note that the sum $\sum_{\De\in \mathcal{S}}$ is only over heavy scalars with $\De>0$. More explicitly, we have
\begin{align}
&y^{\frac{1}{2}}\sum_{\cO \in \widetilde{\mathcal{L}} \cup \mathcal{S}}e^{-2\pi y\De} -y^{\frac{1}{2}}\sum_{\cO\in \widetilde{\mathcal{L}}}\left(-1 +\sum_{m=1}^{\infty} \frac{\zeta(2m)}{\zeta(2m+1)}\frac{2^m\pi^{m+\frac{1}{2}}(-\Delta)^{m}}{m\Gamma(m+\tfrac{1}{2})}y^{-m}\right) \nn \\
-&y^{\frac{1}{2}}\sum_{\cO\in \widetilde{\mathcal{J}}} \left(4\sigma_0(J) +\sum_{m=1}^{\infty}\frac{2\pi^{m+\frac{1}{2}}\sigma_{2m}(J)T_m(\tfrac{-\Delta}{|J|})}{m\Gamma(m+\tfrac{1}{2})|J|^m\zeta(2m+1)}y^{-m}-\sum_{m=1}^{\infty}\frac{2\pi^{m}\sigma_{2m}(J)T_m(\tfrac{-\Delta}{|J|})}{\Gamma(m+1)|J|^m\zeta(2m)}y^{m}\right)  \nn \\
=&\frac{3}{\pi} (\hat Z_p,1)+\sum_{k=1}^{\infty}\mathrm{Re}\left(\de_k y^{\frac{1+z_k}{2}}\right) +\sum_{\cO\in \mathcal{S}}\sum_{n=1}^{\infty} b(n) \sqrt{2\De} e^{-\frac{2\pi n^2\De}{y}}.
\label{eq:crosswithzetastuff}
\end{align}

This equation is one of our main results. It is a crossing equation that relates the large-$y$ and small-$y$ expansion of the scalar part of the primary partition function. It acts on only the heavy scalar operators and the light operators of the 2d CFT. Note that the sums over the heavy scalars are convergent for all CFTs thanks to the $e^{-2\pi y \De}$ and $e^{-\frac{2\pi n^2 \De}{y}}$ factors which suppress the Cardy growth $\sim e^{2\pi\sqrt{\frac{(c-1)\De}{3}}}$. On the right-hand side, there is an interesting term involving the nontrivial zeros of the Riemann zeta function $\sum_{k=1}^{\infty}\mathrm{Re}\left(\de_k y^{\frac{1+z_k}{2}}\right)$, whose coefficient $\de_k$ is theory-dependent. There is also another term $\frac{3}{\pi} (\hat Z_p,1)$ which gives the integral of the (subtracted) partition function over the fundamental domain. In section \ref{sec:numericswithzetas}, we will explore these terms numerically in more detail.

\subsection{Functionals and sum rules}\label{sec:functional_sumrules}

In this section, we will use the crossing equation \eqref{eq:crosswithzetastuff} to derive yet another equation \eqref{eq:maincrossing} that acts only on the scalars of the theory, which allows us to compute numerical upper bounds on the scaling dimension of the leading scalar. The idea is the same as the conformal bootstrap, where one applies linear functionals on the equation to constrain the allowed values of $\De$ of the leading scalar. For this argument to work, we would want the all theory-dependent coefficients in each sum to be positive, but this is not true for the $\de_k$ coefficients in \eqref{eq:crosswithzetastuff}. So, to keep only the sign-definite terms, we should apply functionals that annihilate terms of the form $y^{\frac{1+z_k}{2}}$, where $z_k$'s are the nontrivial zeros of the Riemann zeta functions.

Let us multiply \eqref{eq:crosswithzetastuff} by $y^{-1}$ and redefine $y=t^{-2}$. Then, the equation becomes
\begin{align}
&t\sum_{\cO \in \widetilde{\mathcal{L}} \cup \mathcal{S}}e^{-\frac{2\pi \Delta}{t^2}}-t\sum_{\cO\in \widetilde{\mathcal{L}}}\left(-1 +\sum_{m=1}^{\infty} \frac{\zeta(2m)}{\zeta(2m+1)}\frac{2^m\pi^{m+\frac{1}{2}}(-\Delta)^{m}}{m\Gamma(m+\tfrac{1}{2})}t^{2m}\right) \nn \\
-&t\sum_{\cO\in \widetilde{\mathcal{J}}} \left(4\sigma_0(J) +\sum_{m=1}^{\infty}\frac{2\pi^{m+\frac{1}{2}}\sigma_{2m}(J)T_m(\tfrac{-\Delta}{|J|})}{m\Gamma(m+\tfrac{1}{2})|J|^m\zeta(2m+1)}t^{2m}-\sum_{m=1}^{\infty}\frac{2\pi^{m}\sigma_{2m}(J)T_m(\tfrac{-\Delta}{|J|})}{\Gamma(m+1)|J|^m\zeta(2m)}t^{-2m}\right)  \nn \\
=&\frac{3}{\pi} (\hat Z_p,1)t^2+\sum_{k=1}^{\infty}\mathrm{Re}\left(\de_k t^{1-z_k}\right) +\sum_{\cO\in \mathcal{S}}\sum_{n=1}^{\infty} b(n) \sqrt{2\De} t^2e^{-2\pi n^2\De t^2}.
\label{eq:crosswithzetastuff_tvar}
\end{align}
A family of functionals that can annihilate the $t^{1-z_k}$ terms was given in \cite{Benjamin:2022pnx}. Let us briefly review the definition here. The functionals are defined by
\begin{equation}
    \mathcal{F}^{\varphi}[f(t)] = \int_0^{\infty}\frac{dt}{t}f(t)\(\sum_{n=1}^{\infty}\varphi(n t)\),
\end{equation}
where
\begin{equation}
    \varphi(t) =\sum_{i}^N\alpha_i e^{-\pi r_i^2 t^2}.
\end{equation}
Additionally, $\alpha_i$ and $r_i$ are subject to the following constraints:
\begin{align}\label{eq:functionalcoeff_constraints}
    \sum_{i=1}^N\alpha_i &=0, \nn \\
    \sum_{i=1}^N\alpha_i r_i^{-1} &=0. 
\end{align}
These properties imply that the action of the functional on $t^s$ is given by (see \cite{Benjamin:2022pnx} for derivation)
\begin{equation}
    \mathcal{F}^{\varphi}[t^s] = \frac{\pi^{-\frac{s}{2}}}{2}\zeta(s)\Gamma(\tfrac{s}{2})\sum_i^N\alpha_i r_i^{-s} = \frac{\pi^{-\frac{1-s}{2}}}{2}\zeta(1-s)\Gamma(\tfrac{1-s}{2})\sum_i^N\alpha_i r_i^{-s}.
\end{equation}
In particular, due to the $\zeta(1-s)$ factor, we have $\mathcal{F}^{\varphi}[t^{1-z_k}]=0$.

Next, we can apply the functional to \eqref{eq:crosswithzetastuff_tvar}. As we explain above, the $t^{1-z_k}$ terms vanish. Furthermore, very interestingly the contributions from the light spinning operators also vanish, and we obtain
\begin{align}
    &\sum_i \alpha_i \int d\De\, \rho^{\text{scalars}}_p(\De)\frac{\log\(\frac{1-e^{-2\pi r_i\sqrt{2\De}}}{1-e^{-\frac{2\pi \sqrt{2\De}}{r_i}}}\)}{r_i} +\sum_i \frac{\alpha_i}{2r_i^2} (\hat Z_p,1)-\sum_i\alpha_i\sum_{\cO \in \widetilde{\mathcal{L}}} \frac{\pi\sqrt{2\De}}{r_i^2}=0.
\end{align}
Let us consider the above equation with a single $r$,
\begin{equation}\label{eq:sumrule_singler}
    \int d\De\, \rho^{\text{scalars}}_p(\De)\frac{\log\(\frac{1-e^{-2\pi r\sqrt{2\De}}}{1-e^{-\frac{2\pi \sqrt{2\De}}{r}}}\)}{r} + \frac{1}{2r^2} (\hat Z_p,1) - \sum_{\cO \in \widetilde{\mathcal{L}}}\frac{\pi\sqrt{2\De}}{r^2}.
\end{equation}
Since this equation does not obey the constraints \eqref{eq:functionalcoeff_constraints}, it will not give zero. However, we will get zero when we combine different $r$'s and $\alpha$'s such that they satisfy \eqref{eq:functionalcoeff_constraints}. In order for this to be true, the $r$-dependence of \eqref{eq:sumrule_singler} must be
\begin{equation}\label{eq:sumrule_singler_withcs}
    \int d\De\, \rho^{\text{scalars}}_p(\De)\frac{\log\(\frac{1-e^{-2\pi r\sqrt{2\De}}}{1-e^{-\frac{2\pi \sqrt{2\De}}{r}}}\)}{r} + \frac{1}{2r^2} (\hat Z_p,1) - \sum_{\cO \in \widetilde{\mathcal{L}}}\frac{\pi\sqrt{2\De}}{r^2} = c_0 + c_1 r^{-1},
\end{equation}
for some $r$-independent constants $c_0,c_1$. After multiplying by $r$, the first term on the left-hand side is manifestly antisymmetric in $r \leftrightarrow r^{-1}$. From this, one can show that the constants should be
\begin{equation}
    c_0 = \frac{1}{2}(\hat Z_p,1) - \sum_{\cO \in \widetilde{\mathcal{L}}}\pi\sqrt{2\De},\quad c_1=0.
\end{equation}
Then, \eqref{eq:sumrule_singler_withcs} becomes
\begin{equation}\label{eq:sumrule_singler_final}
    \int d\De\, \rho^{\text{scalars}}_p(\De)\frac{\log\(\frac{1-e^{-2\pi r\sqrt{2\De}}}{1-e^{-\frac{2\pi \sqrt{2\De}}{r}}}\)}{r-r^{-1}} = \frac{1}{2}(\hat Z_p,1) - \sum_{\cO \in \widetilde{\mathcal{L}}}\pi\sqrt{2\De}.
\end{equation}
This gives the crossing equation \eqref{eq:maincrossing} with
\begin{equation}
    \varepsilon=\frac{3}{\pi}(\hat Z_p,1) - \sum_{\cO \in \widetilde{\mathcal{L}}}6\sqrt{2\De}.
    \label{eq:sadfsdf}
\end{equation}
Note of course that the two terms in (\ref{eq:sadfsdf}) correspond to the real and imaginary parts of $\varepsilon$.

Note that the asymptotic Cardy growth of the density of states is $\sim e^{2\pi \sqrt{\frac{(c-1)\De}{3}}}$. On the other hand, at $\De\to \infty$, the factor $\log\(\frac{1-e^{-2\pi r\sqrt{2\De}}}{1-e^{-\frac{2\pi \sqrt{2\De}}{r}}}\)$ goes as $e^{-2\pi \text{min}(r,r^{-1}) \sqrt{2\De}}$. Therefore, in order for the integral in \eqref{eq:sumrule_singler_final} to converge, we must have $c<1+6~\text{min}(r,r^{-1})^2$. The optimal value is $r=1$, where we have $c<7$.

\section{Examples}
\label{sec:examples}

\subsection{Liouville theory}
\label{sec:Liouville}
Let us check \eqref{eq:maincrossing} for Liouville theory. The partition function for Liouville theory is proportional to
\be
Z^{\text{Liouville}}(\tau) \propto \frac{1}{y^{\frac12}|\eta(\tau)|^2}.
\ee
Let us set the constant of proportionality to $\frac1{\sqrt 2}$ which will be convenient later:
\be
Z^{\text{Liouville}}(\tau) \coloneqq \frac{1}{\sqrt 2 y^{\frac12}|\eta(\tau)|^2}.
\ee
The corresponding primary partition function is then simply given by $Z^{\text{Liouville}}_p(\tau)=\frac{1}{\sqrt{2}}$. This gives the density of primary operators to be precisely $\frac{\theta(\Delta)}{\sqrt{\Delta}}$ since
\be
\int_{-\frac{c-1}{12}}^\infty d\Delta \sqrt{y} \rho_p^{\text{Liouville}}(\Delta) e^{-2\pi y \Delta} = \int_0^{\infty} \frac{d\Delta}{\sqrt \Delta} \sqrt y e^{-2\pi y \Delta} = \frac{1}{\sqrt{2}}.
\ee
This means $\varepsilon^{\text{Liouville}} = \frac1{\sqrt 2}$.
And indeed, for all $r$,
\be
\int_0^\infty d\Delta \frac{\log\left(\frac{1-e^{-2\sqrt 2\pi r\sqrt{\Delta }}}{1-e^{-2\sqrt 2 \pi r^{-1} \sqrt{\Delta}}}\right)}{\sqrt{\Delta} (r - r^{-1})} = -\int_0^\infty \frac{du}{\sqrt 2 \pi} \log(1-e^{-u}) =  \frac{\pi}{6\sqrt 2}, 
\ee
which is consistent with (\ref{eq:maincrossing}) with $\varepsilon^{\text{Liouville}} = \frac{1}{\sqrt{2}}$. 

\subsection{MWK}
\label{sec:polestructuremwk}

Let us check the crossing equation we have for MWK \cite{Maloney:2007ud, Keller:2014xba}. There are two cases to check -- a scalar seed and a spinning seed.

\subsubsection{Scalar seed}
First let us do a scalar seed. Following the notation of \cite{Keller:2014xba}, we first consider the Poincar\'{e} sum of a scalar operator of energy $E<0$:
\begin{align}
    Z^{\text{MWK}}(E;\tau,\bar\tau) &= \sum_{\gamma\in\Gamma_\infty\backslash SL(2,\mathbb Z)} \gamma\left( y^{1/2} q^{E/2} \bar{q}^{E/2}\right) \nonumber \\
    &= \sum_{\gamma\in\Gamma_\infty\backslash SL(2,\mathbb Z)} \gamma\left(y^{1/2} e^{-2\pi y E}\right) \nonumber \\
    &\coloneqq  \int_{E}^{\infty} \rho^{\text{scalars}}_{\text{MWK,}~ E}(\Delta) y^{1/2} e^{-2\pi y \Delta} + \sum_{j\neq 0} e^{2\pi i j x} (\ldots).
    \label{eq:mwkexpressionj0}
\end{align}
(For example the vacuum term would be at $E=-\frac{c-1}{12}$.)
In Eqns (71) and (72) of \cite{Keller:2014xba}, the authors explicitly computed the scalar density as:
\be
\rho^{\text{scalars}}_{\text{MWK,}~E}(\Delta ) = \delta(\Delta-E)- \delta(\Delta) + \sum_{m=1}^\infty \frac{2^{4m} \pi^{2m} (-E)^m \zeta(2m)}{\Gamma(2m+1)\zeta(2m+1)}  \Delta^{m-1}.
\label{eq:rhoscalarmwk7172}
\ee 
In Appendix \ref{sec:putcalchere}, we integrate (\ref{eq:rhoscalarmwk7172}) against $f(\Delta, r)$ to get:
\begin{align}
     \int_{E}^{\infty} d\Delta\, \rho^{\text{scalars}}_{\text{MWK,}~E}(\Delta) \left[f(\Delta, r) + f(\Delta, r^{-1})\right] &= 2\pi i \sqrt{-E/2}.
\end{align}
Thus
\be
\varepsilon = 12 i \sqrt{-E/2},
\label{eq:vareps}
\ee
which indeed agrees with \eqref{eq:sadfsdf}. We can obtain (\ref{eq:vareps}) in another way. Formally we would like to define $\varepsilon$ as:
\begin{align}
\varepsilon &\coloneqq \int_{\mathcal{F}} \frac{dx dy}{y^2} \frac{Z(\tau)}{\pi/3} \nonumber \\ 
&= \frac3\pi \int_0^\infty dy y^{-3/2} e^{-2\pi y E}
\end{align}
If $\text{Re}(\Delta) > 0, \text{Re}(s) > -1$ we have
\be
\frac3\pi\int_0^\infty dy y^s e^{-2\pi y \Delta} = \frac3\pi (2\pi\Delta)^{-s-1} \Gamma(s+1)
\label{eq:quickvareps}
\ee
so formally setting $\Delta = E, s=-3/2$, we get
\be
\varepsilon = 12i\sqrt{-E/2}.
\ee 

\subsubsection{Spinning seed}
\label{sec:spinningmwkpoincare}

If we instead consider the Poincar\'{e} sum of an operator with spin $J \neq 0$ and dimension $E$ (with $E<0$ and without loss of generality $J>0$),
\begin{align}
Z^{\text{MWK,}~J}(E,J;\tau,\bar\tau)&= \sum_{\gamma\in \Gamma_\infty \backslash SL(2,\mathbb Z)} \gamma(y^{1/2} q^{\frac{E+J}2} \bar{q}^{\frac{E-J}2}) \nonumber \\
&= \sum_{\gamma\in\Gamma_\infty \backslash SL(2,\mathbb Z)} \gamma\left(y^{1/2} e^{-2\pi y E} e^{2\pi i J x}\right) \nonumber\\
&\coloneqq \int_{E}^{\infty}\rho^{\text{scalars}}_{\text{MWK,}~E,~J}(\Delta) y^{1/2} e^{-2\pi y \Delta} + \sum_{j\neq 0} e^{2\pi i j x} (\ldots).
\label{eq:mwkexpressionspinJ}
\end{align}
We get a scalar density of states of (see (76) of \cite{Keller:2014xba}):
\be
\rho^{\text{scalars}}_{\text{MWK,}~E,~J}(\Delta) = 2\sigma_0(J) \delta(\Delta) + \sum_{m=1}^\infty \frac{2^{3m+1} \pi^{2m} \sigma_{2m}(J) T_m(\frac{-E}J)}{J^m \zeta(2m+1)\Gamma(2m+1)} \Delta^{m-1}
\label{eq:keller76}
\ee
In Appendix \ref{sec:putcalchere} we get
\begin{align}
\int d\Delta \rho^{\text{scalars}}_{\text{MWK,}~E,~J} (\Delta&)\log\left(\frac{1-e^{-2\sqrt 2\pi r \sqrt{\Delta}}}{1-e^{-2\sqrt 2 \pi r^{-1} \sqrt{\Delta}}}\right) = 0.
\end{align}
Note that this is exactly what we expect from the Rankin-Selberg transform, since we took the Poincar\'{e} sum of an expression with no scalar terms.

\subsection{Rademacher Sum} 

We will also check the Rademacher sum of a seed scalar operator, which was computed in \cite{Alday:2019vdr}. The spectrum of a scalar seed with dimension $E<0$ is:
\begin{align}
    \sum_{\gamma\in\Gamma_\infty \backslash SL(2,\mathbb Z),~\text{Rad}} \gamma(y^{1/2} e^{-2\pi y E})  &= \sum_{m=0}^{\infty}\frac{(-2\pi E)^{m+\frac12}}{\Gamma(m+\frac32)} E_{m+1}(\tau,\bar\tau),
    \label{eq:fullexpressionaldaybae}
\end{align}
We put the subscript ``Rad" to emphasize the Rademacher prescription (as opposed to the Poincar\'{e} prescription) is used for the $SL(2,\mathbb Z)$ sum -- see \cite{Alday:2019vdr} for more details. In (\ref{eq:fullexpressionaldaybae}), $E_1$ is defined as a regulated Eisenstein series. Generally the regulated Eisenstein series at $s=1$ is defined up to a constant ambiguity $\mathfrak{c}$: 
\begin{equation}
    E_1(\tau,\bar\tau) \coloneqq y - \frac{3}{\pi} \log y+\mathfrak{c}+ \frac{12}{\pi}\sum_{j=1}^\infty \frac{\cos(2\pi j x)\sigma_1(j) e^{-2\pi j y}}{j}.
    \label{eq:e1defreg}
\end{equation}
However, we claim we should choose $\mathfrak{c}$ as: \begin{equation}\mathfrak c = \frac{3}{\pi}\log(-2\pi E) + \frac9\pi \gamma_E - \frac{36\zeta'(2)}{\pi^3} - \frac6\pi.
\label{eq:mathfrakcdef}
\end{equation} This constant is determined by taking the limit as $m\rightarrow 0$ of the summand of (\ref{eq:fullexpressionaldaybae}), and discarding the divergent piece, i.e. if we expand at small $m$:
\begin{align}
    \frac{(-2\pi E)^{m+\frac12}}{\Gamma(m+\frac32)} &E_{m+1}(\tau,\bar\tau) = \frac{6\sqrt2\sqrt{-E}}{\pi m} \nonumber \\ &+ 2\sqrt2\sqrt{-E} \left(y-\frac{3}{\pi} \log y+\mathfrak{c}+ \frac{12}{\pi}\sum_{j=1}^\infty \frac{\cos(2\pi j x)\sigma_1(j) e^{-2\pi j y}}{j}\right) + \mathcal{O}(m).
    \label{eq:limitnonsingular}
\end{align}
We will see that this particular choice of $\mathfrak{c}$ gives a vanishing $\text{Re}(\varepsilon)$.

The scalar spectrum is \cite{Alday:2019vdr}:
\begin{align}
    \rho^{\text{scalars, Rad}}_E(\Delta) &= \delta(\Delta - E) - \frac{\sqrt{-E}}{\pi\sqrt{\Delta}(\Delta-E)} + \left(\frac{12(\gamma_E + \log (8\pi))}{\pi} + 4 \mathfrak c \right)\frac{\sqrt{-E}}{\sqrt{\Delta}}\nonumber \\ &+ \frac{12 \sqrt{-E}}{\pi \sqrt\Delta} \log \Delta +4\pi\sum_{m=1}^\infty \frac{\zeta(2m+1)}{\zeta(2m+2)} \frac{16^m\pi^{2m}(-E)^{m+\frac12}}{\Gamma(2m+2)} \Delta^{m-\frac12}.
    \label{eq:aldaybaespec}
\end{align}

In Appendix \ref{sec:putcalchere}, we explicitly calculate
\begin{equation}
    \int_{-\frac{c-1}{12}}^\infty d\Delta \rho^{\text{scalars, Rad}}_E(\Delta)\left[f(\Delta, r) + f(\Delta, r^{-1})\right] = 2\pi i \sqrt{-E/2} 
    \label{eq:radeps}
\end{equation} 
independent of $r$ as expected. It is notable that, even though the sum in (\ref{eq:aldaybaespec}) is different from the MWK prescription, the final value of $\varepsilon$ is identical.

\subsection{CFTs at small central charge}
Finally we note that we have taken many exactly solvable CFTs with central charge $c < 7$. Theories we have studied include WZW models, tensor products of minimal models, free boson theories at various points in moduli space, and parafermion theories. We have checked that their scalar Virasoro primaries numerically obey our crossing equation to arbitrarily high precision.

\section{Pole structure in the complex $r$ plane}
\label{sec:poles}

In this section we will show that the pole structure of the crossing equation (\ref{eq:maincrossing}) in the complex $r$ plane is completely fixed by the light scalar spectrum of any CFT. First, let us slightly rewrite (\ref{eq:maincrossing}). Define the new variable
\begin{align}
h &\coloneqq \sqrt{2\Delta}.
\label{eq:betah}
\end{align}
If we take a derivative with respect to $r$ to remove the branch cuts in the log we get
\be
\int_{\mathcal{C}} dh \rho^{\text{scalars}}_p(h) \left[g(h,r) + g(h,r^{-1})\right] = \frac{\varepsilon}{12}
\label{eq:betahsystem}
\ee
where 
\be
g(h,r) \coloneqq \frac{h r^2}{(r^2+1)(e^{2\pi h r}-1)},
\label{eq:newcrosssimpler}
\ee
and we redefine $\rho^{\text{scalars}}_p(h)$ as a sum of delta functions in $h$-space rather than $\Delta$-space (to avoid the measure changing). The contour $\mathcal{C}$ in (\ref{eq:newcrosssimpler}) goes from $\sqrt{\frac{c-1}{6}} i$ to $0i$ and then from $0$ to $\infty$. 

As before, we split the operators into ``light" and ``heavy" where we define light operators to obey:\footnote{For convenience, unlike in earlier sections, here we define light to be $\Delta < 0$, not $\Delta \leq 0$.}
\begin{align}
\Delta_{\text{light}} &\coloneqq \{\Delta < 0\}, ~~~\Delta_{\text{heavy}} \coloneqq \{\Delta \geq 0\}, ~~~ 
\end{align}
or equivalently 
\begin{align}
h_{\text{light}} & \in i\mathbb R, ~~~h_{\text{heavy}} \in\mathbb R. ~~~ 
\end{align}
There are only a finite number of light operators; we call their values of $h$:
\be
i Q_1, i Q_2, \ldots i Q_N,
\ee
with $0 < Q_1 < Q_2 < \ldots < Q_N = \sqrt{\frac{c-1}6}$. (The last $Q_N = \sqrt{\frac{c-1}6}$ because it comes from the vacuum which has $\Delta = -\frac{c-1}{12}$.)

From the explicit expression for $g(h,r)$, we see that
\be
\int_{i Q_N}^{i 0} dh \rho^{\text{light}}(h) g(h, r)
\ee
has poles in $r$ at 
\be
r = \frac{k}{Q_n}, ~~~~k\in \mathbb Z, ~~n = 1, 2, \ldots, N,
\ee
and similarly of course
\be
\int_{i Q_N}^{i 0} dh \rho^{\text{light}}(h) g(h, r^{-1})
\ee
has poles in $r$ at 
\be
r = \frac{Q_n}{k}, ~~~~k\in \mathbb Z, ~~n = 1, 2, \ldots, N. 
\ee
This implies that for the heavy part,
\be
\int_0^\infty dh \rho^{\text{heavy}}(h) \left[g(h,r) + g(h,r^{-1})\right]
\ee
has poles with opposite residue at the exact same locations,
\be
r = \frac{k}{Q_n}, \frac{Q_n}{k}, ~~~~k\in \mathbb Z, ~~n = 1, 2, \ldots, N.
\ee
From the results in Sec. \ref{sec:polestructuremwk} we see that for the MWK partition function,
\be
\int_0^\infty dh \rho^{\text{heavy}}(h) g(h, r)
\label{eq:rhoheavypoles}
\ee
has poles at $r = \frac{Q_n}{k}$. This implies that all CFT partition functions have this pole structure. The reason is we can consider the difference $Z_{\text{CFT}} - Z_{\text{MWK}}$, where we subtract out the MWK sum of the light spectrum. This function has no exponential growth so its heavy spectrum converges against $g(h,r)$ for all $r$ so it has no poles. Thus $Z_{\text{CFT}}$ has the same pole structure.

This implies that for any CFT,
\be
\int_{iQ_N}^{i0} dh \rho^{\text{light}}(h) g(h,r^{-1}) + \int_0^\infty dh \rho^{\text{heavy}}(h) g(h,r)
\label{eq:analytic}
\ee
is an analytic function of $r$ in the region $\text{Re}(r) > 0$ (no poles in the right half plane).

For MWK, (\ref{eq:analytic}) is equal to
\be
\frac{i\sqrt{-E/2}}{r^2+1}.
\label{eq:polestuffmwk}
\ee
We can do a similar calculation for Liouville theory. In the $h$-variables, the spectrum is given by: $\rho^{\text{light}}(h) = 0, \rho^{\text{heavy}}(h) = \sqrt 2$. Thus we get 
\begin{equation}
    \int_0^\infty dh \sqrt 2 g(h, r) = \frac1{12\sqrt2 (r^2+1)}.
    \label{eq:polestuffliouville}
\end{equation}
As expected, neither expression (\ref{eq:polestuffmwk}) nor (\ref{eq:polestuffliouville}) has any poles for $\text{Re}(r) > 0$.

Since the integral of $Z_{\text{CFT}} - Z_{\text{MWK}}$ explicitly converges to a real number, and in Sec. \ref{sec:polestructuremwk}, we computed explicitly $\varepsilon$ for MWK, this implies the imaginary part of $\varepsilon$ is fixed for any CFT in terms of its light scalar operators. In particular, we again find
\be
\text{Im}(\varepsilon) = 6\sum_{n=1}^N Q_n,
\label{eq:varepsilonmwkbla}
\ee
which agrees with \eqref{eq:sadfsdf}. Indeed, all examples we check in Sec. \ref{sec:examples} obey (\ref{eq:varepsilonmwkbla}).

We also note that if the central charge obeys $c \leq \frac52$, then $\text{Re}(\varepsilon) > 0$. Since $c<7$ the integral 
\be
\varepsilon = \int_{i Q}^{i 0} dh \rho(h) \frac{12h}{e^{2\pi h}-1} + \int_{0}^{\infty} dh \rho(h) \frac{12h}{e^{2\pi h}-1} 
\ee
converges.
The second integral is manifestly real and positive. The real part of the first integral goes as
\be
\sum_{n=1}^N 6 Q_n \cot(\pi Q_n).
\ee
Since $Q_n \leq \frac 12$ (by assumption of $c \leq \frac52$), this is positive as well. This means that, unlike for Narain CFTs, there is no sense that averaging over all CFTs at a fixed $c$ has vanishing $\text{Re}(\varepsilon)$.\footnote{As we showed in Sec. \ref{sec:polestructuremwk}, MWK partition function itself has vanishing $\text{Re}(\varepsilon)$, even for $c\leq \frac52$. The reason this is not a contradiction is because the scalar spectrum of MWK has a negative delta function at $h=0$, which precisely cancels the positive contribution from the other scalars. For Narain CFTs, \cite{Benjamin:2023chz} discussed $\varepsilon$ in the context of an abstract distance measure. It would be interesting if using the techniques here, we can generalize this to all CFTs, though with analytic continuation to define $\varepsilon$ it is unclear if it would obey some of the axioms of a good distance measure.}

It would be extremely interesting if one could use the pole structure derived in this section to extend the convergence of our main crossing equation (\ref{eq:maincrossing}) beyond central charge $7$.
 
\section{Numerics on crossing equation with zeta zeros}
\label{sec:numericswithzetas}

\subsection{Modular integral of Virasoro primaries}\label{sec:numerics_withzetazero_primaries}
Using \eqref{eq:maincrossing}, we can compute a regulated modular integral $\varepsilon$ for CFTs with $c<7$. We now show that using the other crossing equation \eqref{eq:schematicwow} (or more explicitly \eqref{eq:crosswithzetastuff}), we can explicitly compute a regulated modular integral for any partition function, not just those with $c<7$.

As an example let us consider the $(E_8)_1$ WZW model. This theory has central charge $c=8$, and the partition function is given by
\begin{align}
    Z^{(E_8)_1}(\tau,\bar\tau) &= j(\tau)^{1/3}\bar{j}(\bar\tau)^{1/3},
\end{align}
where $j(\tau)$ is the unique meromorphic modular function with simple pole at $\tau=i\infty$ and $q$-expansion:
\begin{equation}
    j(\tau) = \frac1q + 744 + O(q).
\end{equation}

We can define $Z_p^{\text{scalars}, ~(E_8)_1}(\tau,\bar\tau)$, and its inverse Laplace transform $\rho^{\text{scalars}, ~(E_8)_1}$ as
\begin{align}
    Z_p^{\text{scalars},~ (E_8)_1}(y) &= \int_{-1/2}^{1/2} dx y^{\frac12} |\eta(\tau)|^2 Z^{(E_8)_1}(\tau,\bar\tau) \nonumber \\
    &= y^{\frac12}\left(e^{\frac{7\pi y}{6}} + 61009 e^{-\frac{17\pi y}{6}}+15015625 e^{-\frac{41\pi y}{6}}+ \ldots\right), \nonumber \\
    \rho_p^{\text{Vir scalars},~(E_8)_1~}(\Delta) &= \delta(\Delta + \frac{7}{12}) + 61009 ~\delta(\Delta - \frac{17}{12}) + 15015625 ~\delta(\Delta-\frac{41}{12}) + \ldots.
\label{eq:scalarse8vir}
\end{align}
Since $c>7$, if we plug the spectrum (\ref{eq:scalarse8vir}) into (\ref{eq:maincrossing}), the equation diverges for every value of $r$. Here, however, we show that we can still extract $\varepsilon$ using the crossing equation (\ref{eq:crosswithzetastuff}).

\begin{figure}[H]
\begin{center}
  \includegraphics[width=0.9\linewidth]{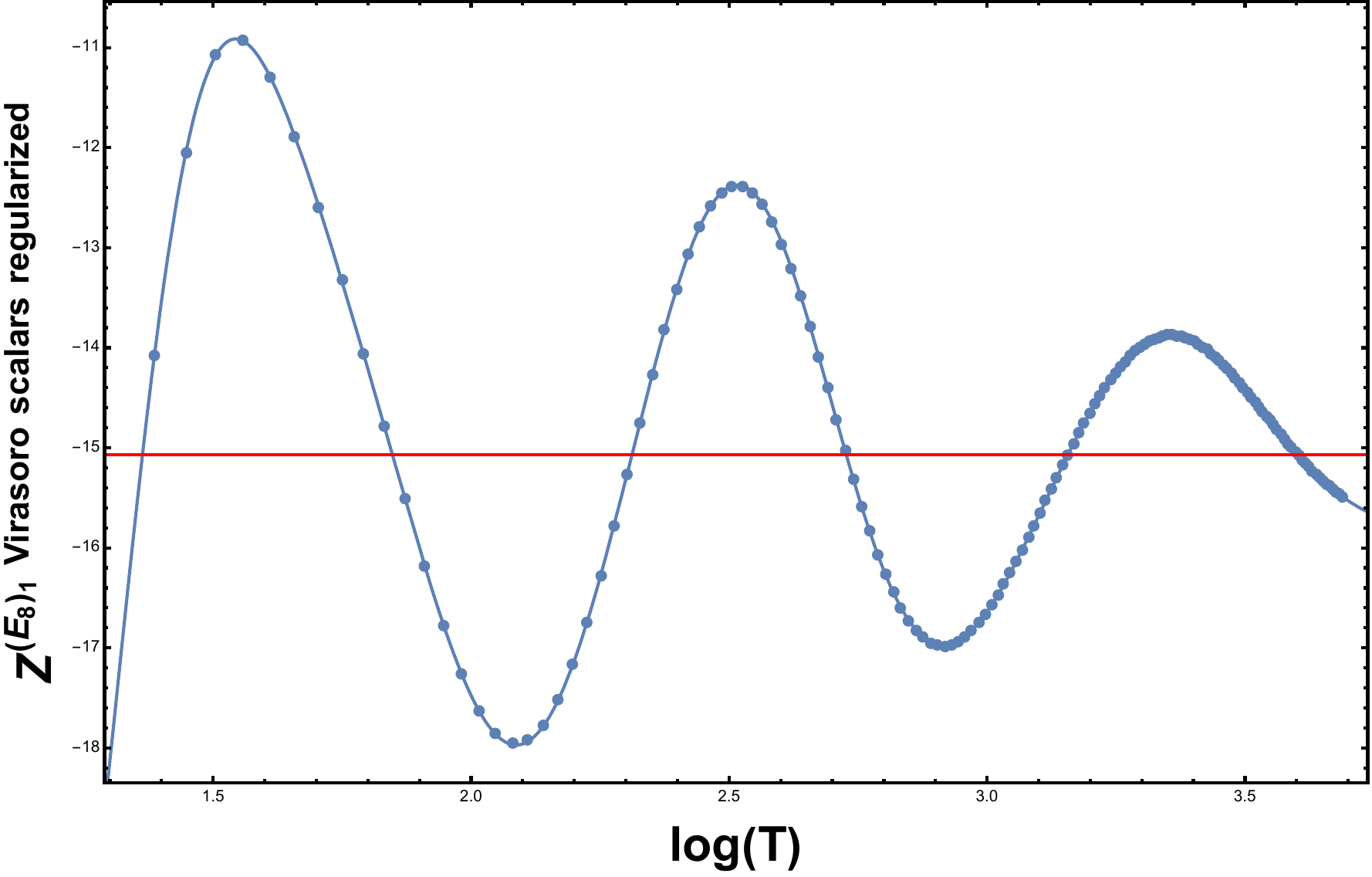}
  \end{center}
  \caption{A plot of the scalar partition function of the Virasoro primary $(E_8)_1$ WZW partition function with the MWK spectrum subtracted out. The resulting function is fitted to the large $T$ prediction in (\ref{eq:crosswithzetastuff}), namely $\text{Re}(\varepsilon) - \sum_{k=1}^\infty \text{Re}\left(\delta_k T^{-\frac{1+z_k}2}\right)$ (with the sum in $k$ cut off at $5$) to high precision. The fit gives $\text{Re}(\varepsilon) \approx -15.07$.}
  \label{fig:ZetaE8}
\end{figure}

In Fig. \ref{fig:ZetaE8} we plot the scalar partition function (\ref{eq:scalarse8vir}) at large $T \coloneqq y^{-1}$ (equivalently small $y$) with the MWK sum of the light operator subtracted out. (More precisely, we plot the LHS of \eqref{eq:crosswithzetastuff}.) From (\ref{eq:crosswithzetastuff}) we see this should be well-approximated at large $T$ by a constant $\varepsilon$ plus an infinite sum of decaying oscillations, whose frequency is controlled by the imaginary parts of the nontrivial zeros of the zeta function. By fitting this, we estimate $\text{Re}(\varepsilon) \approx -15.07$.

This leads to a definite prediction that
\begin{equation}
    \frac{3}{\pi} \int_{\mathcal{F}} \frac{dxdy}{y^2}\sqrt{y}|\eta(\tau)|^2 (j(\tau)^{1/3}\bar{j}(\bar\tau)^{1/3}) \approx -15.07 + 12i\sqrt{\frac{8-1}{24}}.
\end{equation}

Note that the plot in Fig. \ref{fig:ZetaE8} (as well as the three other plots below) is also numerical evidence for the assumption \eqref{eq:ZpEs_overlap_assumption} made in the derivation of the crossing equation. The fact that the plotted function goes to a constant at large $T$ can be directly translated to the $O(\Delta^{-1/2})$ behavior in the RHS of \eqref{eq:ZpEs_overlap_assumption}.\footnote{For the three other plots in Sec. \ref{sec:modularzetas0}, the behavior becomes $O(\Delta^{-1})$ since there is no additional $y^{1/2}$ power.}

\subsection{Modular integral of full partition function}
\label{sec:modularzetas0}

We can do a similar calculation for the modular integral of a partition function as opposed to a ``reduced" partition function (i.e. without an overall $y^{1/2} |\eta(\tau)|^2$ multiplying it). 

We first need to redo the calculation in Sec. \ref{sec:crossing_derivation} by doing harmonic analysis without an overall power $y^{1/2}$ multiplying the partition function. For the sum rules on scalars analogous to (\ref{eq:maincrossing}), this was done in \cite{Benjamin:2023nts}. If the partition function is written as (note that there is no $y^{1/2}$ in (\ref{eq:partitionfunctionnoy}))
\begin{equation}
    Z(\tau,\bar\tau) = \int_{-\frac{c}{12}}^\infty \rho^{\text{scalars}}(\Delta) e^{-2\pi y \Delta} +\sum_{j\neq 0} e^{2\pi i j x} (\ldots)\, ,
    \label{eq:partitionfunctionnoy}
\end{equation}
then
\begin{align}
    \varepsilon &\coloneqq \frac3\pi \int_{\mathcal{F}} \frac{dx dy}{y^2} Z(\tau,\bar\tau) \nonumber \\ &= \frac{12\sqrt2}{\pi(r-r^{-1})} \int_{-\frac{c}{12}}^\infty d\Delta \rho^{\text{scalars}}(\Delta) \sqrt\Delta\sum_{m=1}^\infty \frac{K_1(2\pi m r^{-1} \sqrt{2\Delta}) - K_1(2\pi m r \sqrt{2\Delta})}{m}.
    \label{eq:varepsnoystuff}
\end{align} 
Just as in (\ref{eq:maincrossing}), the RHS is $r$-independent. This expression has the largest convergence as $r$ approaches $1$. As $r$ approaches $1$, (\ref{eq:varepsnoystuff}) converges for $c<6$. Note that in (\ref{eq:varepsnoystuff}) (and throughout this subsection), $\Delta$ is now the conformal weight with $\frac{c}{12}$ subtracted, rather than $\frac{c-1}{12}$, since we are counting all states, not just Virasoro primaries.

There similarly is an analogous equation for the other crossing equation \eqref{eq:crosswithzetastuff}. Just as before, we need to subtract out the Poincar\'{e} sum of the light states. In App. \ref{app:generalypower}, we write down general expressions for Poincar\'{e} sums of light operators with an arbitrary power of $y^{w/2}$ in front. Here we are interested in the case of $w=0$. More precisely we will need to do sums of the form 
\begin{align}
    \sum_{\gamma \in \Gamma_\infty \backslash SL(2,\mathbb Z)} \gamma(q^{\frac{E+J}2}\bar{q}^{\frac{E-J}2}) &= Z^{\text{MWK scalar states}}_{E,J}(y) + \sum_{j\neq 0} e^{2\pi i j x} (\ldots)
    \nonumber \\&= \int_E^\infty d\Delta \rho^{\text{MWK scalar states}}_{E,~J}(\Delta) e^{-2\pi y \Delta} + \sum_{j\neq 0} e^{2\pi i j x}(\ldots).
    \label{eq:modifiedmwk}
\end{align}
for some $E<0$ and spin $J$. (The sum in (\ref{eq:modifiedmwk}) differs form (\ref{eq:mwkexpressionj0}) and (\ref{eq:mwkexpressionspinJ}) because of no $y^{1/2}$ in the sum.) In App. \ref{app:generalypower} we will show that if $J=0$, then\footnote{The $\frac{1}{\Delta}$ term in $\rho^{\text{MWK scalar states}}_{E,J=0}(\Delta)$ should be understood as the ``plus-distribution," whose integral against a test function $f(\Delta)$ is given by $\int d\Delta\,\frac{f(\Delta)-f(0)}{\Delta}$.}
\begin{align}
    Z^{\text{MWK scalar states}}_{E,J=0}(y) &= e^{-2\pi y E} - 6E\left(-1+3\gamma_E-\frac{12\zeta'(2)}{\pi^2}+\log\left(-\frac{\pi E}{2y}\right)\right)\nonumber \\&~~+\sqrt\pi\sum_{m=2}^\infty \frac{(-2\pi E)^m y^{1-m} \Gamma(m-\frac12)\zeta(2m-1)}{m \Gamma(m)^2\zeta(2m)}, \nonumber \\
    \rho^{\text{MWK scalar states}}_{E,~J=0}(\Delta) &= 
    \delta(\Delta - E) - 6E\left(-1+4\gamma_E - \frac{12\zeta'(2)}{\pi^2} +2\log(\pi) +  \log\left(-E\right)\right) \delta(\Delta)
    \nonumber \\& + \left( - \frac{6E}{\Delta} + \frac{1}{2\sqrt\pi} \sum_{m=2}^\infty \frac{(m-1)(-4\pi^2 E)^m \Gamma(m-\frac12)\zeta(2m-1)}{m\Gamma(m)^3 \zeta(2m)}\Delta^{m-2}\right)\theta(\Delta),
\label{eq:statesdensityspinJ0}
\end{align}
and if $J \neq 0$
\begin{align}
   &Z^{\text{MWK scalar states}}_{E,J}(y)  = \sum_{a,s=0}^\infty\frac{(-2\pi E)^s(-1)^a \pi^{2a+\frac12}\sigma_{4a+2s-1}(|J|)\Gamma(a+s-\frac12)y^{1-2a-s}}{\Gamma(s+1)\Gamma(a+1)\Gamma(2a+s)\zeta(4a+2s)|J|^{2a+2s-1}}, \nonumber \\
   &\rho^{\text{MWK scalar states}}_{E,J}(\Delta) =  - \frac{12 E}{|J|}\sigma_1(|J|)\delta(\Delta) \nonumber \\ &~~~+ \frac{1}{2\sqrt\pi} \sum_{a,s=0}^{\infty}\!\!{}^\prime \frac{(-4\pi^2 E)^s(-1)^a (4\pi^4)^a\sigma_{4a+2s-1}(|J|)\Gamma(a+s-\frac12)(2a+s-1)}{\Gamma(s+1)\Gamma(a+1)\Gamma(2a+s)^2\zeta(4a+2s)|J|^{2a+2s-1}} \Delta^{2a+s-2}\theta(\Delta),
\label{eq:statesdensityspinJnon0}
\end{align} 
where $\sum_{a,s=0}^{\infty}\!\!{}^\prime$ means we omit the $a=s=0$ and $a=0,s=1$ terms in the sum. 

By plugging the scalar densities (\ref{eq:statesdensityspinJ0}) and (\ref{eq:statesdensityspinJnon0}) into (\ref{eq:varepsnoystuff}) we get $\varepsilon$ for the subtracted MWK sums. In particular we get:
\begin{align}
    \varepsilon^{\text{MWK},E,J=0} &= 1-6\pi E i, \nonumber \\
    \varepsilon^{\text{MWK},E,J \neq 0} &= -12\sigma_1(|J|).
    \label{eq:epsilonsmwkw0}
\end{align}
Surprisingly, we get a nonvanishing $\varepsilon$ for a spinning seed, unlike in Sec. \ref{sec:spinningmwkpoincare}.\footnote{We can understand the imaginary part of the first equation in (\ref{eq:epsilonsmwkw0}), $\text{Im}(\varepsilon) = -6\pi E$, by taking the $s\rightarrow -2$ limit of (\ref{eq:quickvareps}) (with $\Delta=E)$ and extracting the nonsingular piece.} 

At small $y$, we get a crossing equation of the form
\begin{align}
    &Z^{\text{scalars}}(y) - \sum_{E\leq0} Z^{\text{MWK scalar states}}_{E,J}(y) = \varepsilon - \sum_{E\leq0} \varepsilon^{\text{MWK}, E,J} + \sum_{k=1}^\infty \text{Re}(\delta_k y^{\frac{1+z_k}2}) + \sum_{\substack{E\leq 0 \\ J=0}}\left(1-e^{-2\pi E y}\right)\nonumber\\
    &+ \sum_{\substack{E\leq 0 \\ J\neq0}}\sum_{a,s=0}^\infty\!\!{}^\prime  \frac{(-1)^{a+s} (-2\pi E)^s \pi^{2a-1} \Gamma(\frac32-2a-s)\Gamma(-\frac12+a+s)\sigma_{4a+2s-1}(|J|)}{|J|^{2a+2s-1}\Gamma(a+1)\Gamma(s+1)\zeta(4a+2s-1)}y^{2a+s} +O(e^{-\#/y}),
    \label{eq:fullcrossingnovirjuststates}
\end{align}
where again the primed sum indicates we remove the $a=s=0$ and $a=0,s=1$ terms. (See \eqref{eq:crosiingwithzeta_states} for a full expression.) 

If there are no states with $E\leq 0, J\neq 0$ (which is automatic from unitarity if $c<12$), (\ref{eq:fullcrossingnovirjuststates}) simplifies substantially. It becomes
\begin{align}
    Z^{\text{scalars}}(y) - \sum_{E \leq 0} \left(Z^{\text{MWK scalar states}}_{E,J=0}(y) - e^{-2\pi E y} - 6\pi E i\right) &= \varepsilon - \sum_{k=1}^\infty \text{Re}(\delta_k y^{\frac{1+z_k}2}) + O(e^{-\#/y}).
    \label{eq:subtractionwhennospinning}
\end{align}
Even for states with $E\leq 0, J \neq 0$, we still can efficiently compute $\varepsilon$ numerically. We will now consider four examples.

\begin{figure}[H]
\begin{center}
  \includegraphics[width=0.9\linewidth]{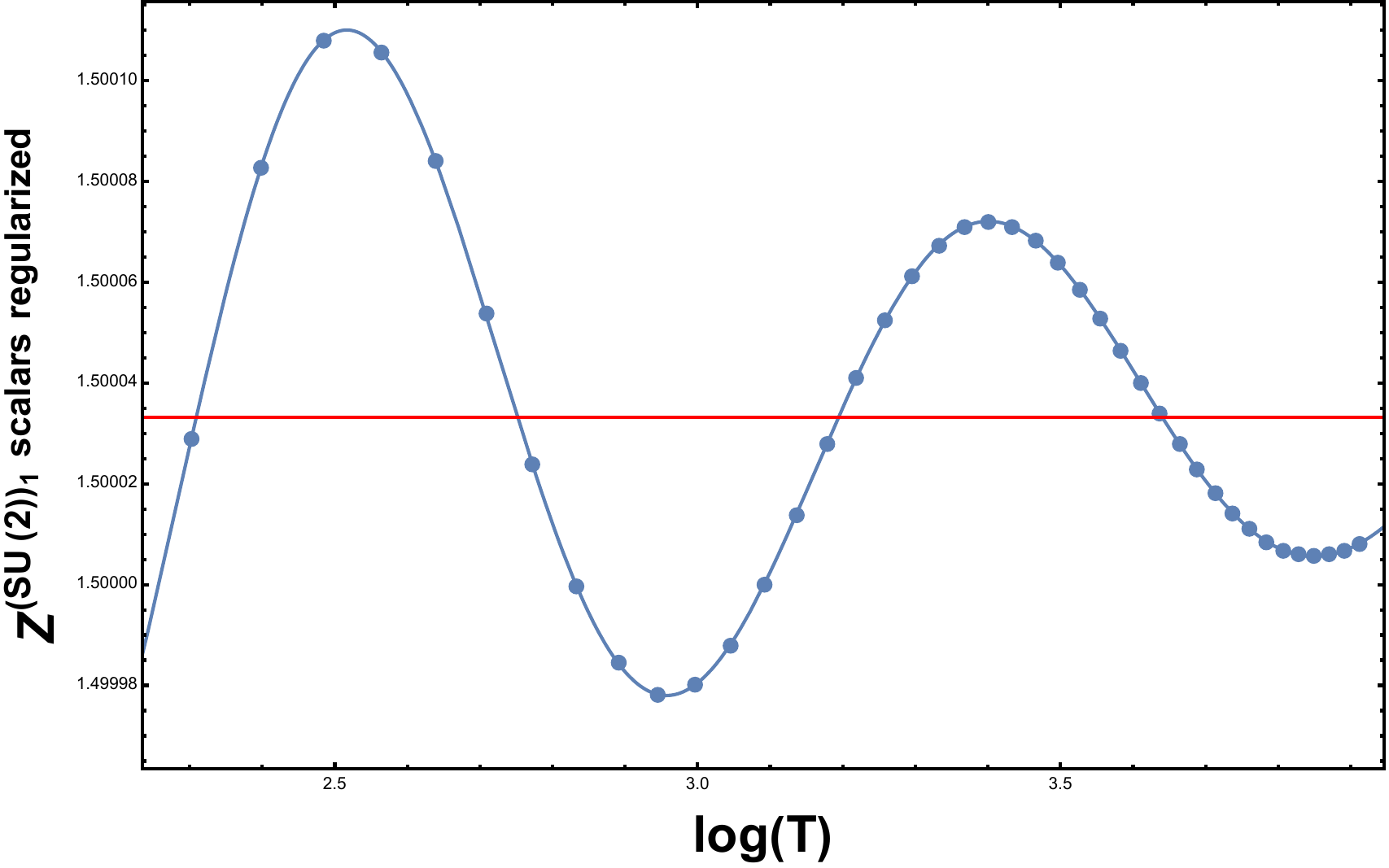}
  \end{center}
  \caption{A plot of the scalar partition function of the $(SU(2))_1$ WZW partition function with the MWK spectrum subtracted out as in the real part of (\ref{eq:subtractionwhennospinning}). The resulting function is fitted to the large $T$ prediction in (\ref{eq:crosswithzetastuff}), namely $\text{Re}(\varepsilon) - \sum_{k=1}^\infty \text{Re}\left(\delta_k T^{-\frac{1+z_k}2}\right)$ (with the sum in $k$ cut off at $5$) to high precision. The fit gives $\text{Re}(\varepsilon) \approx 1.50003$. This is consistent with Fig. 7 of \cite{Baccianti:2025gll} (note the factor of $\frac\pi3$ that differs from the convention in \cite{Baccianti:2025gll}).} 
  \label{fig:SU2Zeta}
\end{figure}

The first example we consider is the $SU(2)_1$ WZW model, which has $c=1$. The only light state we need to subtract out is the vacuum, so we take the scalar partition function and subtract (\ref{eq:statesdensityspinJ0}) with $E=-\frac1{12}$ and plot at large $T \coloneqq y^{-1}$. We plot the result in Fig. \ref{fig:SU2Zeta}, which gives as an explicit estimate for $\varepsilon$. More precisely,
\begin{equation}
     \frac{3}{\pi} \int_{\mathcal{F}} \frac{dxdy}{y^2} Z^{SU(2)_1}(\tau,\bar\tau) \approx 1.50003 + \frac{\pi i}{2}.
\end{equation}  
This is consistent with the expression we get for $\varepsilon$ from (\ref{eq:varepsnoystuff}), as well as a calculation done in \cite{Baccianti:2025gll}.

\begin{figure}[H]
\begin{center}
  \includegraphics[width=0.9\linewidth]{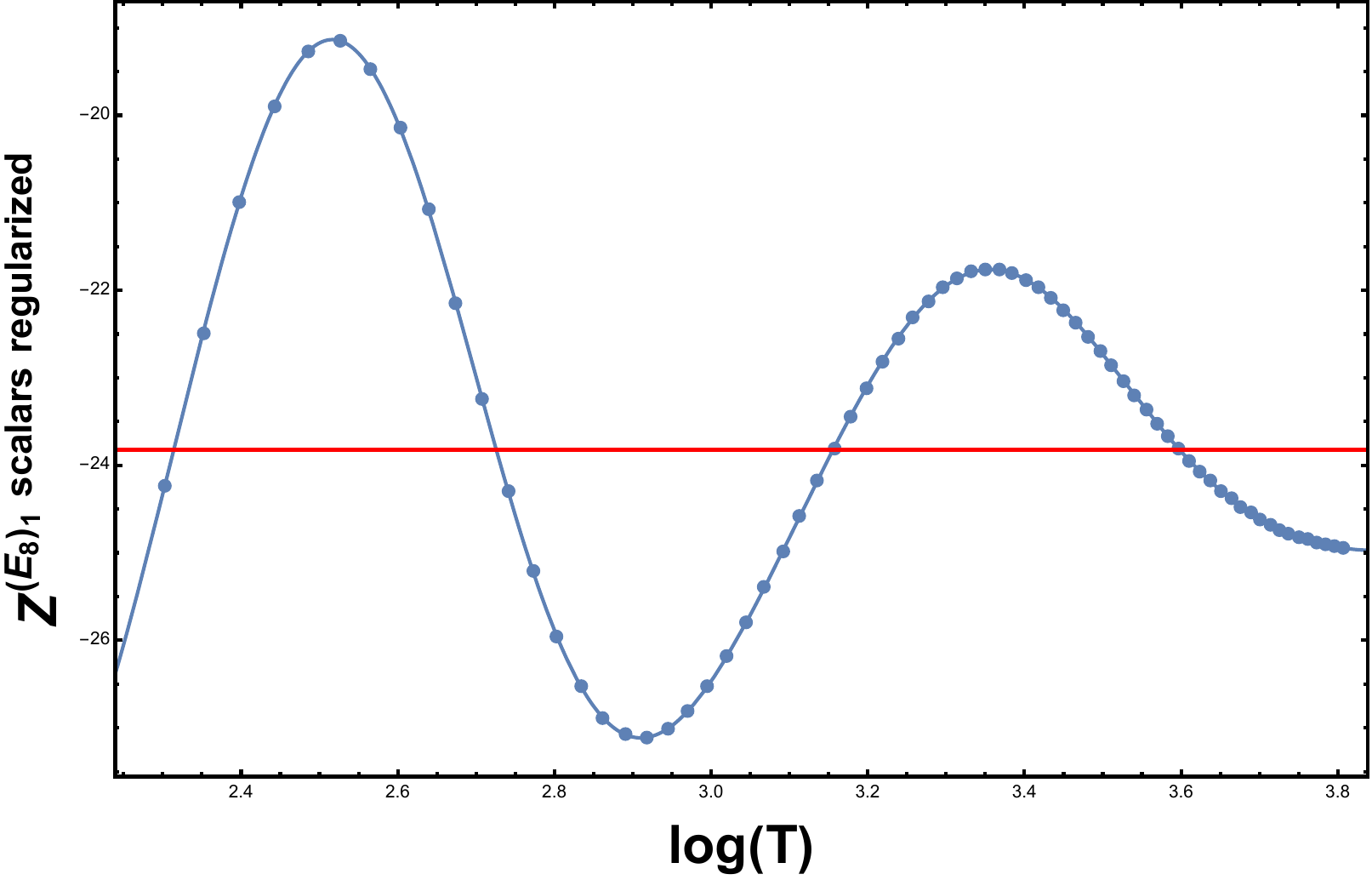}
  \end{center}
  \caption{A plot of the scalar partition function of the $(E_8)_1$ WZW partition function with the MWK spectrum subtracted out as in the real part of (\ref{eq:subtractionwhennospinning}). The resulting function is fitted to the large $T$ prediction in (\ref{eq:crosswithzetastuff}), namely $\text{Re}(\varepsilon) - \sum_{k=1}^\infty \text{Re}\left(\delta_k T^{-\frac{1+z_k}2}\right)$ (with the sum in $k$ cut off at $5$) to high precision. The fit gives $\text{Re}(\varepsilon) \approx -23.823$.}
  \label{fig:ZetaE8states}
\end{figure}  

The second example we will do is the $(E_8)_1$ WZW model again. Compared to the previous case in Sec. \ref{sec:numerics_withzetazero_primaries}, now we will integrate the entire partition function, as opposed to the partition function with $y^{1/2} |\eta(\tau)|^2$ multiplying it. We plot the result in Fig. \ref{fig:ZetaE8states}, which gives
\begin{equation}
     \frac{3}{\pi} \int_{\mathcal{F}} \frac{dxdy}{y^2} j(\tau)^{1/3}\bar{j}(\bar\tau)^{1/3} \approx -23.823 + 4\pi i.
\end{equation}
Since the central charge $c>6$, the expression (\ref{eq:varepsnoystuff}) does not converge, but we nevertheless are able to compute $\varepsilon$.

\begin{figure}[H]
\begin{center}
  \includegraphics[width=0.9\linewidth]{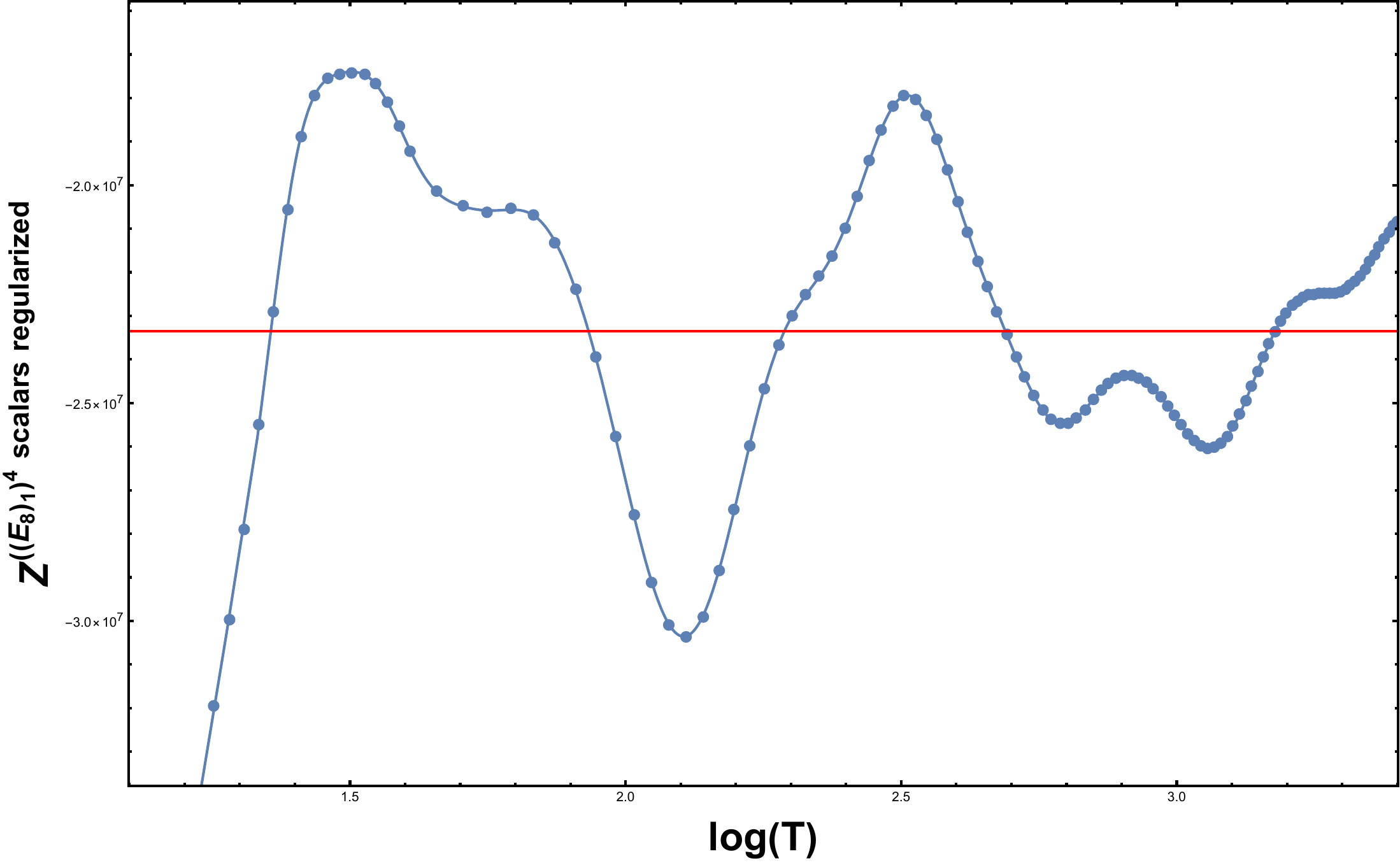}
  \end{center}
  \caption{A plot of the scalar partition function of the $((E_8)_1)^4$ WZW partition function with the light spectrum subtracted out as in (\ref{eq:fig4monstrosity}). Unlike previous examples, this theory has states with $E<0$ of nonzero spin. The resulting function is fitted to the large $T$ prediction in (\ref{eq:crosswithzetastuff}), namely $\text{Re}(\varepsilon) - \sum_{k=1}^\infty \text{Re}\left(\delta_k T^{-\frac{1+z_k}2}\right)$ (with the sum in $k$ cut off at $20$) to high precision. The fit gives $\text{Re}(\varepsilon) \approx -2.3 \times 10^7$.}
  \label{fig:ZetaE84states}
\end{figure} 

The third example we do is the $((E_8)_1)^4$ WZW model. This is a theory at central charge $c=32$. Unlike the previous examples, this theory has states with $E\leq 0$ and $J\neq 0$. In particular there are 1984 states at $|J|=1, E=-5/3$, and 771040 states at $|J|=2, E=-2/3$. In Fig. \ref{fig:ZetaE84states}, we plot
\begin{align}
   &Z^{\text{scalars}}(y) - \sum_{\substack{E \leq 0 \\ J=0}} \left(Z^{\text{MWK scalar states}}_{E,J=0}(y) - e^{-2\pi E y}\right) \nonumber \\ &- \sum_{\substack{E \leq 0 \\ J\neq0}}\Bigg(Z^{\text{MWK scalar states}}_{E, J}(y) + 12\sigma_1(|J|) \nonumber \\ &~~~~~~~~~~~~~~+\sum_{a,s=0}^\infty\!\!{}^\prime  \frac{(-1)^{a+s} (-2\pi E)^s \pi^{2a-1} \Gamma(\frac32-2a-s)\Gamma(a+s-\frac12)\sigma_{4a+2s-1}(|J|)}{|J|^{2a+2s-1}\Gamma(a+1)\Gamma(s+1)\zeta(4a+2s-1)}y^{2a+s}\Bigg).
   \label{eq:fig4monstrosity}
\end{align}
By \eqref{eq:fullcrossingnovirjuststates}, the above quantity should equal to a constant plus a sum of oscillations controlled by zeros of the zeta function, and we find the fit gives a constant of about $-2.3\times 10^7$. In total we get
\begin{equation}
     \frac{3}{\pi} \int_{\mathcal{F}} \frac{dxdy}{y^2} j(\tau)^{4/3}\bar{j}(\bar\tau)^{4/3} \approx -2.3\times10^7 + 20\pi i.
\end{equation}

Our fourth and final example is the integral of a meromorphic modular function. For every given $j \in \mathbb{Z}_{\geq 0}$, let us consider the unique meromorphic modular function with a pole of order $j$ at $\tau=i\infty$ (and nowhere else) that obeys
\begin{equation}
    Z^{\text{meromorphic}}_j(\tau) \coloneqq \frac1{q^j} + \mathcal{O}(q^1).
\end{equation}
The MWK sum of $1/q^j$ can be seen from e.g. (\ref{eq:statesdensityspinJnon0}) with $-E=J=j$, and it gives
\begin{equation}
    \sum_{\gamma\in \Gamma_\infty \backslash SL(2,\mathbb Z)} \gamma(q^{-j}) = Z_j^{\text{meromorphic}}(\tau) + 12 \sigma_1(j).
\end{equation}
We can also verify that if $-E=J$, then for all $y$,
\begin{equation}
 \sum_{a,s=0}^\infty\!\!{}^\prime  \frac{(-1)^{a+s} (2\pi |J|)^s \pi^{2a-1} \Gamma(\frac32-2a-s)\Gamma(-\frac12+a+s)\sigma_{4a+2s-1}(|J|)}{|J|^{2a+2s-1}\Gamma(a+1)\Gamma(s+1)\zeta(4a+2s-1)}y^{2a+s} = 0.
 \end{equation}
 Therefore, plugging into (\ref{eq:fullcrossingnovirjuststates}), we get:
 \begin{equation}
     \frac{3}{\pi} \int_{\mathcal{F}} \frac{dxdy}{y^2} Z^{\text{meromorphic}}_j(\tau) = -24\sigma_1(j),
 \end{equation}
 which precisely agrees with e.g. \cite{Baccianti:2025gll, Lerche:1987qk}.

\section{Numerics on the scalar gap}
\label{sec:numerics}

We now present our numerical results for bounds on the scalar gap of general 2d CFTs that contain solely Virasoro symmetry without any extended chiral algebra. By applying a generalization of the prescription used in \cite{Hellerman:2009bu}, we use our new crossing equation to extract rigorous bounds for various values of $c<7$.

Given the positivity of $\rho^{\text{scalars}}_p(\Delta)$ and the fact that derivatives of the (manifestly $r$-independent) RHS of our crossing equation (\ref{eq:maincrossing}) must vanish, a natural and computationally convenient basis of functionals for our bootstrap calculation is  

\be \text{vac}^{(n)}(1) + \sum_{\Delta} (\partial_r)^n F( \Delta, r)|_{r=1} = 0, ~~n\geq 1,
\label{eq:functionalbasis}
\ee
where the sum is over the set of conformal dimensions of scalar Virasoro primary operators shifted by $-\frac{c-1}{12}$. For the expression above, we define
\be
F(\Delta, r) \coloneqq f(\Delta, r) + f(\Delta, r^{-1})
\ee
and 
\be \text{vac}^{(n)}(r) \coloneqq  
(\partial_r)^n 
\left[
F(-\frac{c-1}{12}, r) + F(-\frac{c-25}{12}, r)
\right].
\label{eq:vacuum}
\ee

Note that the second contributing term in (\ref{eq:vacuum}) follows from the Virasoro vacuum character's null state structure. We assume that there are no additional spin 1 currents present, otherwise the equation would be slightly modified (for $c<\frac32$ this is automatic from unitarity).

Taking combinations of the functionals in (\ref{eq:functionalbasis}) with different orders of derivatives gives a positive sum rule for the shifted operators. It turns out that using consecutive values of $n$ in the functional basis does not lead to independent equations, and therefore the functionals take the following generic form:
\be
\sum_{n=2,4,\ldots,n_{\mathrm{max}}} \alpha_n \text{vac}^{(n)}(1) + \sum_{n=2,4,\ldots,n_{\mathrm{max}}} \alpha_n \sum_{\Delta}(\partial_r)^{n}F(\Delta,r)|_{r=1} =0.
\label{eq:sumrule}
\ee
Any given sum rule can thus be characterized by the highest derivative taken, $n_{\mathrm{max}}$.

The general argument used to derive a bound in the scalar gap from (\ref{eq:sumrule}) is straightforward. We begin by proposing that the gap in the shifted spectrum of primary operators is some given test value which we denote by $\Delta^*$. If it can be shown that there exists at least one set of $\{\alpha_n\} \in \mathbb{R}^{n_{\text{max}}/2}$ such that 

\begin{align}
&\sum_{n=2,4,\ldots,n_{\mathrm{max}}} \alpha_n \text{vac}^{(n)}(1) >0, \nonumber \\
&\sum_{n=2,4,\ldots,n_{\mathrm{max}}} \alpha_n(\partial_r)^{n}F(\Delta,r)|_{r=1} \geq 0,\qquad \Delta \geq \Delta^*,
\label{eq:constraintsforarg}
\end{align}
we are then able to refute the hypothesized spectrum because (\ref{eq:constraintsforarg}) is in contradiction with the positivity of our crossing equation's density of states. This means that $\Delta^*$ is then a rigorous upper bound on the shifted scalar gap.

For large numbers of derivatives, determining whether such a combination of functionals exists for a given $\Delta^*$ becomes highly nontrivial. To complete this objective, we used the semidefinite program solver {\tt SDPB}
\cite{Simmons-Duffin:2015qma,Landry:2019qug}. 

In practice, the second constraint in (\ref{eq:constraintsforarg}) is evaluated for discrete set of sample values $\Delta_1,\Delta_2,\ldots,\Delta_{M} \geq \Delta^*$, with the program seeking a set $\{\alpha_n\} \in \mathbb{R}^{n_{\text{max}}/2}$ for which the functional is positive at each $\Delta_m$. The corresponding continuous condition for all $\Delta \geq \Delta^*$ is then verified by manually inspecting a plot of the suggested functional. If this reveals any negative parts for $\Delta \geq \Delta^*$, we add more sample points in the negative regions and rerun the solver. This process is concluded once either the resulting functional is completely positive or {\tt SDPB} declares that no such functional exists. 

For ease of comparison with results from other works, the calculated bounds we quote here are for the scalar spectrum without the $-\frac{c-1}{12}$ offset. For this non-shifted spectrum, we use the notation $\Delta^{(n_{\mathrm{max}})}_{\text{scalar gap}}$ to represent the strongest upper bound obtained characterized by the highest derivative order $n_{\mathrm{max}}$.

\begin{table}[t]
\begin{center}
    \begin{tabular}{| c | c | c | c| c|  }
    \hline
    $c$ &$\Delta_{\text{scalar gap}}^{(10)}$&$\Delta_{\text{scalar gap}}^{(20)}$ &$\Delta_{\text{scalar gap}}^{(30)}$ &$\Delta_{\text{scalar gap}}^{(40)}$   \\ \hline
    1.0001 & 0.50600 & 0.50008 & 0.50002 & 0.50002 \\
    1.1 & 0.52612 & 0.51706 & 0.51687 & 0.51684 \\
    1.2 & 0.54771 & 0.53466 & 0.53420 & 0.53408  \\
    1.3 & 0.57069 & 0.55292 & 0.55203 & 0.57043  \\ 
    1.4 & 0.59494 & 0.57189 & 0.57043 & 0.56991  \\ 
    1.5 & 0.62046 & 0.59162 & 0.58944 & 0.58857  \\
    2.0 & 0.768 & 0.703 & 0.695 & 0.691 \\
    2.5 & 0.956 & 0.843 & 0.825 & 0.814  \\
    3.0 & 1.203 & 1.022 & 0.989 & 0.966  \\
    3.5 & 1.543 & 1.264 & 1.209 & 1.167 \\
    4.0 & 2.043 & 1.611 & 1.520 & 1.450 \\
    4.5 & 2.836 & 2.149 & 1.999 & 1.880 \\ 
    5.0 & 4.247 & 3.086 & 2.827 & 2.613 \\
    5.5 & 7.212 & 5.014 & 4.512 & 4.096 \\
    6.0 & 15.508 & 10.306 & 9.096 & 8.0875 \\

    6.5 & 59.617 & 37.878 & 32.790 & 28.515 \\

    \hline
    \end{tabular}
    \caption{Upper bounds on the scalar gap of general $2d$ CFTs for various $c<7$ obtained from different instances of (\ref{eq:sumrule}) characterized by $n_{\mathrm{max}} = 10,20, 30 ,40$.}
    \label{tab:boundstable}
\end{center} 
\end{table}

\begin{figure}[t]
\begin{center}
  \includegraphics[width=1.0\linewidth]{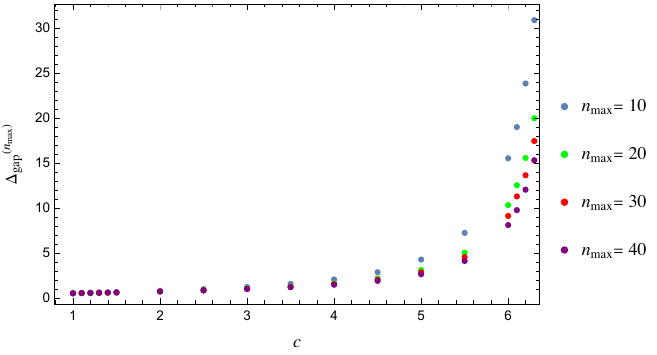}
  \end{center}
  \caption{Plot of a bound on the scalar gap for general $2d$ CFTs at various $c < 7$. The colors blue, green, red, and purple represent the bound we get at $10, 20, \cdots, 40$ derivatives respectively. See Table \ref{tab:boundstable} for the numerical data.}
  \label{fig:plotu1}
\end{figure}

Using the method described above, we have computed $\Delta^{(n_{\mathrm{max}})}_{\text{scalar gap}}$ for $n_{\mathrm{max}}=10, 20, \cdots, 40$ at various values of central charge  $c < 7$. Our bounds are summarized in Table \ref{tab:boundstable} and plotted in Fig. \ref{fig:plotu1}.

 For $c\in [1,4]$, the results appear to corroborate bound $c/6+1/3$, matching result for $c$ very close to 1 from \cite{Collier:2016cls}, but bounds become generally weaker than those found in \cite{Collier:2016cls} for larger $c$, with bounds diverging as we approach $c$ to $7$ since sum rule only converges for $c<7$.
 
 The scalar gap bound diverged at $c\geq 25$ in \cite{Collier:2016cls} due to a spurious solution to the modular bootstrap equations with no scalar primaries:
 \begin{equation}
     Z^{\text{spurious}}(\tau,\bar\tau) = \frac{1}{y^{1/2}|\eta(\tau)|^2}\left((j(\tau)-744+\bar{j}(\bar\tau)-744\right).
\label{eq:spurious}
 \end{equation}
It would be interesting if we could somehow do an analytic continuation in $r$ to extend our crossing equation to a convergent expression for all $r$. If so, it is conceivable we could get a nontrivial scalar gap for all $c$, because by definition, any functional obtained from manipulations of (\ref{eq:maincrossing}) vanishes on spectrum of (\ref{eq:spurious}). As is, of course,  our bounds in Table \ref{tab:boundstable} are still substantially weaker than the scalar gap bounds in \cite{Collier:2016cls}.

\section{Discussion}
\label{sec:discuss}

There are a number of interesting future directions we would like to point out.

It has been long-known that the zeros of the Riemann zeta function exhibit chaotic properties, in particular are well-approximated by a GUE matrix distribution \cite{Montgomery,Odylzko,SarnakZeev}. It is extremely tempting to look at our crossing equation (\ref{eq:schematicwow}) which we reproduce below
\begin{align}
    Z^{\text{scalars}}_p(y) - Z^{\text{gravity}}_{j=0}(y) &= \varepsilon + \sum_{k=1}^\infty \text{Re}\left(\delta_k y^{\frac{1+z_k}2}\right) \nonumber \\& + (\text{perturbative terms fixed by light spectrum})(y) + O(e^{-\#/y}),
\label{eq:schematicwow2}
\end{align}
and speculate if the distribution of zeros of the Riemann zeta function have anything to do with distribution of primary operators in generic CFTs. One obvious complication is that (\ref{eq:schematicwow}) is true for all CFTs, both rational and irrational (indeed the examples we looked at in Sec. \ref{sec:numericswithzetas} were all RCFTs), so the chaotic properties in the scalar sector would somehow be encoded the residues $\delta_k$. It would also be natural to take (\ref{eq:schematicwow}) and analytically continue $y \rightarrow y + i t$ and compute the spectral form factor\footnote{See e.g. \cite{Cotler:2016fpe, Dyer:2016pou, Benjamin:2018kre} for discussion on the SFF in chaotic and integrable theories. See also \cite{Haehl:2023mhf, Haehl:2023tkr,DiUbaldo:2023qli, Boruch:2025ilr} for related discussion on quantum chaos and harmonic analysis.}.

Besides in the context of 2d CFTs, modular functions also naturally show up in other areas of physics. The recent papers \cite{Dorigoni:2023nhc, Godet:2025bju} discuss the connection between scalar sectors of other modular functions in physics and zeros the Riemann zeta function. It would be interesting to make some of the connections to this paper more precise. 

Finally, in this paper we have focused on scalars in a 2d CFT. This is partially due to the fact that for other spinning sectors, one has to take into account the Maass cusp forms, a much more intricate set of eigenfunctions of the Laplacians. Nonetheless, we can write down formal expressions for crossing equations in other spin sectors of a CFT.

We can consider the spin $j$, $j > 0$ sector of a 2d CFT's reduced partition function $\hat Z(\tau,\bar\tau)$, and using the same logic as in Sec. \ref{sec:Derivecross}, subtract out the Poincar\'{e} sum of all light states. This time, we will keep the spin $j$ component of the Poincar\'{e} sum, which we denote as $Z^{\text{gravity}}_j(y)$. After deforming the contour over $s$, we get:
\begin{align}
&Z^{\text{spin~} j}_p(y) - Z^{\text{gravity}}_j(y) = \sum_{k=1}^\infty \text{Re}\left(\delta_{k} \frac{\sigma_{z_k}(j) K_{\frac{z_k}{2}}(2\pi j y) y^{\frac{1}2}}{\Lambda\left(\frac{z_k+1}{2}\right) j^{\frac{z_k}2}}\right)  \nonumber \\ & ~~~+  \frac1{2\pi i} \int d\De\, (\rho^{\text{scalars}}_p(\De)-\rho^{\text{gravity}}_{j=0}(\De))\sqrt{2y\Delta} \sum_{n=1}^\infty n \mu(n) \int_{\gamma-i\infty}^{\gamma+i\infty} ds (2\Delta n^2)^{- s} \frac{\sigma_{2s-1}(j) K_{s-\frac12}(2\pi j y)}{j^{s-\frac12}} \nonumber \\
&~~~+ \sum_{n=1}^{\infty}  \frac{a_j^{(n,+)}(\hat Z, \nu_n^{+})}{2(\nu_n^{+},\nu_n^{+})}  \sqrt y K_{i R_n^+}(2\pi j y)+ \sum_{n=1}^{\infty}  \frac{a_j^{(n,-)}(\hat Z, \nu_n^{-})}{2i(\nu_n^{-},\nu_n^{-})}  \sqrt y K_{i R_n^-}(2\pi j y),
\label{eq:spinjcrossingCleanedUp}
\end{align}  
where $\mu(n)$ is the M\"obius function, and $\Lambda(s) \coloneqq \pi^{-s} \Gamma(s) \zeta(2s)$. In the last line, we describe the contribution coming from the Maass cusp forms, following the notation in \cite{Benjamin:2021ygh}. The $s$-integral in the second line can be done exactly, as well as the $\De$-integral of the $\rho^{\text{gravity}}$ term. We find
\begin{align}
&Z^{\text{spin~} j}(y) - Z^{\text{gravity}}_j(y) \nonumber \\
&= \sum_{k=1}^\infty \text{Re}\left(\delta_{k} \frac{\sigma_{z_k}(j) K_{\frac{z_k}{2}}(2\pi j y) y^{\frac{1}2}}{\Lambda\left(\frac{z_k+1}{2}\right) j^{\frac{z_k}2}}\right) + \frac{\sqrt y}2 \sum_{\Delta \in \mathcal{S}} \sum_{n=1}^\infty \mu(n) \sum_{k | j} e^{-\pi y j \left(\frac{2\Delta n^2 j}{k^{2}} + \frac{k^2}{2\Delta n^2 j}\right)} \nonumber \\
&-\sqrt{y}\sum_{\cO \in \widetilde{\mathcal{L}}}\sum_{m=1}^{\infty}\frac{2^m \pi ^{2 m+\frac{1}{2}}  (-E)^m \sigma _{2 m}(j) K_m(2 \pi  j y)}{\zeta (2 m+1) \Gamma \left(m+\frac{1}{2}\right) \Gamma (m+1)j^m} \nonumber \\
&-\sqrt{y}\sum_{\cO \in \widetilde{\mathcal{J}}} \sum_{m=1}^{\infty} \frac{2 \pi ^{2 m+\frac{1}{2}}  \sigma _{2 m}(j) \sigma _{2 m}(|J|) T_m\left(\frac{-E}{|J|}\right) K_m(2 \pi  j y)}{\zeta (2 m) \zeta (2 m+1) \Gamma \left(m+\frac{1}{2}\right) \Gamma (m+1)j^m |J|^m}\nonumber \\
& + \sum_{n=1}^{\infty}  \frac{a_j^{(n,+)}(\hat Z, \nu_n^{+})}{2(\nu_n^{+},\nu_n^{+})}  \sqrt y K_{i R_n^+}(2\pi j y)+ \sum_{n=1}^{\infty}  \frac{a_j^{(n,-)}(\hat Z, \nu_n^{-})}{2i(\nu_n^{-},\nu_n^{-})}  \sqrt y K_{i R_n^-}(2\pi j y).
\label{eq:spinjcrossingCleanedUpv2}
\end{align}  
As an example of (\ref{eq:spinjcrossingCleanedUpv2}), let us consider a $c=1$ free boson at radius $r$. For these theories, the primary counting partition function has vanishing overlap with the Maass cusp forms, and also vanishing residues $\delta_k$ \cite{Benjamin:2021ygh ,Benjamin:2022pnx}. The scalar states come from primary operators with vanishing momentum or vanishing winding modes, which have energy $\frac{m^2}{2r^2}$ or $\frac{m^2 r^2}2$. Finally, $Z^{\text{gravity}}_j(y)$ vanishes for these theories since the Poincar\'e sum of the vacuum becomes the Eisenstein series $E_{1/2}$ which is identically zero. Thus (\ref{eq:spinjcrossingCleanedUpv2}) simplifies to
\begin{align}
Z^{\text{spin~} j}_{\text{free boson~}r}(y) &= \frac{\sqrt y}2 \sum_{\Delta \in \mathcal{S}} \sum_{n=1}^\infty \mu(n) \sum_{k | j} e^{-\pi y j \left(\frac{2\Delta n^2 j}{k^{2}} + \frac{k^2}{2\Delta n^2 j}\right)} \nonumber \\
&= \sqrt y\sum_{n=1}^\infty \mu(n) \sum_{k | j} \sum_{m=1}^\infty \left( e^{-\pi y \left(\frac{m^2 n^2 j^2}{k^{2} r^2} + \frac{r^2k^2}{m^2 n^2}\right)} + e^{-\pi y \left(\frac{m^2 r^2 n^2 j^2}{k^{2}} + \frac{k^2}{m^2 r^2 n^2}\right)} \right) \nonumber \\
&=  \sqrt y\sum_{n=1}^\infty \sum_{k | j}\left( e^{-\pi y \left(\frac{n^2 j^2}{k^{2} r^2} + \frac{r^2k^2}{n^2}\right)} + e^{-\pi y \left(\frac{r^2 n^2 j^2}{k^{2}} + \frac{k^2}{r^2 n^2}\right)} \right)  \sum_{m| n}  \mu(m) \nonumber \\
&= \sqrt y\sum_{k | j} \left( e^{-\pi y \left(\frac{j^2}{k^{2} r^2} + r^2k^2\right)} + e^{-\pi y \left(\frac{r^2 j^2}{k^{2}} + \frac{k^2}{r^2}\right)} \right) \nonumber \\
&= 2\sqrt y\sum_{k | j} e^{-\pi y\left(\frac{k^2}{r^2} + \frac{j^2}{k^2} r^2\right)}.
\label{eq:spinjcrossingCleanedUpv3}
\end{align}  
(In going from the third to fourth line we used the fact that $\sum_{m|n} \mu(m) =\delta_{n,1}$.) We can indeed recognize the last line of (\ref{eq:spinjcrossingCleanedUpv3}) as all primary operators in the $c=1$ free boson at radius $r$ of spin $j$. 

It would be very interesting if we could use the more general structure of (\ref{eq:spinjcrossingCleanedUpv2}) to understand the chaotic nature of primary partition functions at spin $j$, as well as relate to the results in e.g. \cite{Haehl:2023mhf, Haehl:2023tkr,DiUbaldo:2023qli, Boruch:2025ilr} on RMT-universality from harmonic analysis. 

\section*{Acknowledgments}
We are grateful to Scott Collier, Lorenz Eberhardt, Yuya Kusuki, Wei Li, Alex Maloney, Dalimil Maz\'{a}\v{c}, Hirosi Ooguri, Sridip Pal, and Eric Perlmutter for very helpful discussions. We thank Scott Collier, Lorenz Eberhardt, and Tom Hartman for comments on a draft. NB and TR are supported by the US Department of Energy, Office of Science, Office of High Energy Physics, under Award Number DE-SC0025704. CHC is supported by a Kadanoff fellowship at the University of Chicago.

\appendix
\section{Details of the crossing equation derivation}\label{app:crossing}
\subsection{Expressions of  $Z^{\mathrm{MWK}}$}\label{app:MWK_details}
We will be particularly interested in the scalar part of $Z^{\mathrm{MWK}}_{\cO}$ (note that $\cO$ can be either scalar or spinning operator). In \cite{Keller:2014xba}, these were computed explicitly. For scalar $\cO$, we have
\begin{align}\label{eq:ZMWK_scalar}
\int_{-\frac{1}{2}}^{\frac{1}{2}} dx\, Z^{\mathrm{MWK}}_{E,J=0}(\tau,\bar \tau) =& y^{\frac{1}{2}}\left(e^{-2\pi y E} -1 +\sum_{m=1}^{\infty} \frac{\zeta(2m)}{\zeta(2m+1)}\frac{2^m\pi^{m+\frac{1}{2}}(-E)^{m}}{m\Gamma(m+\tfrac{1}{2})}y^{-m}\right) \nn \\
=&y^{\frac{1}{2}}  \int_0^{\infty} d\De\,  \rho^{\text{scalars}}_{\text{MWK},E,J=0}(\De)\, e^{-2\pi y \De},
\end{align}
where
\be\label{eq:rhoMWK_scalarseed}
\rho^{\text{scalars}}_{\text{MWK},E,J=0}(\De)= \de(\De-E)-\de(\De) + \sum_{m=1}^{\infty} \frac{\zeta(2m)}{\zeta(2m+1)}\frac{2^{2m}\pi^{2m+\frac{1}{2}}(-E)^{m}}{\Gamma(m+1)\Gamma(m+\tfrac{1}{2})}\De^{m-1}.
\ee

For spinning $\cO$, we have
\begin{align}\label{eq:ZMWK_spinning}
\int_{-\frac{1}{2}}^{\frac{1}{2}} dx\, Z^{\mathrm{MWK}}_{E,J}(\tau,\bar \tau)  =& y^{\frac{1}{2}} \left(2\sigma_0(J) +\sum_{m=1}^{\infty}\frac{2\pi^{m+\frac{1}{2}}\sigma_{2m}(J)T_m(\tfrac{-E}{|J|})}{m\Gamma(m+\tfrac{1}{2})|J|^m\zeta(2m+1)}y^{-m}\right) \nn \\
=&y^{\frac{1}{2}}\int_0^{\infty} d\De\,  \rho^{\text{scalars}}_{\text{MWK},E,J}(\De)\, e^{-2\pi y \De},
\end{align}
where
\be\label{eq:rhoMWK_spinningseed}
\rho^{\text{scalars}}_{\text{MWK},E,J}(\De)= 2\sigma_0(J)\de(\De)+ \sum_{m=1}^{\infty}\frac{2^{m+1}\pi^{2m+\frac{1}{2}}\sigma_{2m}(J)T_m(\tfrac{-E}{|J|})}{\Gamma(m+1)\Gamma(m+\tfrac{1}{2})|J|^m\zeta(2m+1)} \De^{m-1}.
\ee
$\sigma_{2m}(J)$ is the divisor sigma function, and $T_m$ is the Chebyshev polynomial of the first kind.

\subsection{Properties of $(\hat{Z}_p,E_{s})$}\label{app:ZpEs_poles}

In this appendix, we study the properties of $(\hat{Z}_p,E_{s})$, in particular its pole structure in the complex $s$ plane for $\mathrm{Re}(s) \geq \frac{1}{2}$. We will mainly follow the argument in \cite{Benjamin:2021ygh}. The idea is to take the inverse Laplace transform of \eqref{eq:Vircross_expr0}, and we get
\begin{align}\label{eq:discreteness_eq0}
    &\rho^{\text{scalars}}_p(\De) - \sum_{\cO\in \widetilde{\mathcal{L}}\cup \widetilde{\mathcal{J}}}\rho^{\text{scalars}}_{\text{MWK},E,J}(\De) \nonumber \\
    &=\frac{3}{\pi}(\hat{Z}_p,1)\sqrt{\frac{2}{\De}} + \frac{1}{2\pi i}\int_{\textrm{Re}(s)=\frac{1}{2}}ds\, \frac{2^{\frac{1}{2}-s}\pi^{-s}\Gamma(s)\zeta(2s)\De^{-\frac{1}{2}-s}}{\Gamma(\tfrac{1}{2}-s)\Gamma(s-\frac{1}{2})\zeta(2s-1)}(\hat Z_p,E_{1-s}),
\end{align}
where we have used the functional identity \eqref{eq:ZpEs_identity}:
\begin{equation}\label{eq:ZpEs_identity_inappendix}
    (\hat Z_p,E_s) = \frac{\Gamma(s)\zeta(2s)}{\pi^{\frac{1}{2}}\Gamma(s-\frac{1}{2})\zeta(2s-1)}(\hat Z_p,E_{1-s}).
\end{equation}

Let $N^{\text{scalars}}(\Delta)$ be the number of scalar primaries with scaling dimension below $\De$. Then, by integrating \eqref{eq:discreteness_eq0} we have
\begin{align}
    &N^{\text{scalars}}(\Delta) - \sum_{\cO\in \widetilde{\mathcal{L}}\cup \widetilde{\mathcal{J}}}N^{\text{scalars}}_{\text{MWK},E,J}(\De) \nn \\
    &=\frac{6\sqrt{2}}{\pi}(\hat{Z}_p,1)\sqrt{\De} + \frac{1}{2\pi i}\int_{\textrm{Re}(s)=\frac{1}{2}}ds\, \frac{2^{\frac{1}{2}-s}\pi^{-s}\Gamma(s)\zeta(2s)\De^{\frac{1}{2}-s}}{\Gamma(\tfrac{3}{2}-s)\Gamma(s-\frac{1}{2})\zeta(2s-1)}(\hat Z_p,E_{1-s}),
\end{align}
where $N^{\text{scalars}}_{\text{MWK},E,J}(\De) = \int_{0}^{\De}d\De' \rho^{\text{scalars}}_{\text{MWK},E,J}(\De')$. We can consider the $\De \to 0$ limit of the above equation, where $N^{\text{scalars}}(\De) = 0$ does not contribute due to discreteness of the spectrum. (Note that contribution from the light scalars are exactly canceled by the seed contribution in $N^{\text{scalars}}_{\text{MWK},E,J=0}(\De)$.) In the $\De \to 0$ limit, we can deform the contour of the $s$-integral to the left and pick up the poles in $\text{Re}(s)<1/2$. These poles should reproduce $\frac{6\sqrt{2}}{\pi}(\hat{Z}_p,1)\sqrt{\De}$ and the $N^{\text{scalars}}_{\text{MWK},E,J}(\De)$ contributions. First, it's clear that the integrand has a pole at $s=0$ due to the $\Gamma(s)$ factor in the numerator. In particular, we have
\begin{equation}
    \text{Res}_{s=0}\frac{2^{\frac{1}{2}-s}\pi^{-s}\Gamma(s)\zeta(2s)\De^{\frac{1}{2}-s}}{\Gamma(\tfrac{3}{2}-s)\Gamma(s-\frac{1}{2})\zeta(2s-1)}(\hat Z_p,E_{1-s}) = -\frac{6\sqrt{2}}{\pi}(\hat Z_p,1)\sqrt{\De},
\end{equation}
which is equivalent to the fact that $(\hat Z_p,E_s)$ has a pole at $s=0$ with residue $\frac{3}{\pi}(\hat Z_p,1)$.

Additionally, to reproduce the contributions from $N^{\text{scalars}}_{\text{MWK},E,J}(\De)$, there should also be poles at $s=\frac{1}{2}-m$, $m=0,1,2,\ldots$. The residues of $(\hat Z_p,E_{1-s})$ at these poles are given by
\begin{align}\label{eq:overlap_polefromMWK}
    \text{Res}_{s=\frac{1}{2}}(\hat Z_p,E_{1-s}) &= -\sum_{\cO\in \widetilde{\mathcal{L}}}1 +\sum_{\cO\in\widetilde{\mathcal{J}}}2\sigma_0(J), \nn \\
    \text{Res}_{s=\frac{1}{2}-m}(\hat Z_p,E_{1-s}) &= -\sum_{\cO\in \widetilde{\mathcal{L}}}\frac{(-2\pi E)^m}{\Gamma(1+m)} - \sum_{\cO\in\widetilde{\mathcal{J}}}\frac{2\pi^m\sigma_{2m}(J)T_m(\tfrac{-E}{|J|})}{|J|^m\Gamma(m+1)\zeta(2m)}, \quad m=1,2,\ldots
\end{align}
where the pole at $s=\frac{1}{2}$ should be understood as being slightly to the left of the contour.

Now, we can use the functional identity \eqref{eq:ZpEs_identity_inappendix} to obtain the pole structure of $(\hat Z_p,E_s)$ for $\text{Re}(s)>\frac{1}{2}$. The above argument implies that $(\hat Z_p,E_s)$ has poles at $s=\frac{1}{2}+m$ for $m=0,1,2,3,\ldots$, where the $s=\frac{1}{2}$ pole should be slightly to the right of the $\text{Re}(s)>\frac{1}{2}$ contour. Moreover, the prefactor in \eqref{eq:ZpEs_identity_inappendix} has poles at $\frac{1+z_k}{2},\frac{1+z_k^*}{2}$ due to the $\zeta(2s-1)$ in the denominator, and hence $(\hat Z_p,E_s)$ can also have poles there. In summary, for $\text{Re}(s)>\frac{1}{2}$, $(\hat Z_p,E_s)$ has poles at $s=\frac{1}{2}+m, m=0,1,2,\ldots$ and $s=\frac{1+z_k}{2},\frac{1+z_k^*}{2}$.

Finally, let us comment on how the pole structure is consistent with the assumption \eqref{eq:ZpEs_overlap_assumption_afteridentity}. The idea is that the poles at $s=\frac{3}{2},\frac{5}{2},\ldots$ exactly capture the divergences of the integral \eqref{eq:ZpEs_overlap_assumption_afteridentity} at small $\De$. More precisely, near $s=\frac{1}{2}+m$ the integral has a pole due to the $\De^{m-1}$ term in $\rho^{\text{scalars}}_{\text{MWK}}$. In particular, using the identity
\begin{equation}
    \De^{m-1+\frac{1}{2}-s}\theta(\De) = \frac{1}{\frac{1}{2}+m-s}\delta(\De) +O((\tfrac{1}{2}+m-s)^0),
\end{equation}
the integral \eqref{eq:ZpEs_overlap_assumption_afteridentity} near $s=\frac{1}{2}+m$ can be written as 
\begin{align}
    &(\hat{Z}_p,E_s) = \frac{1}{s-\frac{1}{2}-m} \bigg(\sum_{\cO\in \widetilde{\mathcal{L}}}\frac{(-2\pi E)^{m}}{\Gamma(m+1)}+\sum_{\cO\in \widetilde{\mathcal{J}}}\frac{2\pi^{m}\sigma_{2m}(J)T_m(\tfrac{-E}{|J|})}{\Gamma(m+1)|J|^m\zeta(2m)} \bigg) +O((\tfrac{1}{2}+m-s)^0),
\end{align}
which indeed agrees with the previous prediction for the residue \eqref{eq:overlap_polefromMWK}.

\section{Explicitly evaluating crossing equation for Poincar\'{e} and Rademacher sums}
\label{sec:putcalchere}

In this appendix we check the calculations explicitly done in Section \ref{sec:examples}.

First we check the MWK sum for a scalar seed. This means we explicitly integrate (\ref{eq:rhoscalarmwk7172}) against $f(\Delta, r)$ using the fact that
\begin{align}
\int_0^\infty d\Delta \Delta^{m-1} \log(1-e^{-2\sqrt 2 \pi r\sqrt{\Delta}}) &= r^{-2m} \pi^{-2m} 2^{1-3m} \int_0^\infty du u^{2m-1} \log(1-e^{-u}) \nonumber \\
&= -r^{-2m} \pi^{-2m} 2^{1-3m} \Gamma(2m) \zeta(2m+1),
\end{align} 
which implies
\begin{align}
    \int_{E}^{\infty} &d\Delta\, \rho^{\text{scalars}}_{\text{MWK,}~E}(\Delta) f(\Delta, r) \nonumber \\ &= \frac{1}{r - r^{-1}} \left[ \log(1-e^{-2\sqrt 2\pi i r\sqrt{-E}})-\log(2\pi r\sqrt{2 \epsilon}) -\sum_{m=1}^\infty \frac{(-2E)^m \zeta(2m)}{m r^{2m}} \right],
\end{align}
where $\epsilon$ is a regulator (coming from the $\delta(\Delta)$ term) that will cancel when we add $f(\Delta, r^{-1})$.
Now we use the fact that
\begin{equation}
     -\sum_{m=1}^\infty \frac{(-2E)^m \zeta(2m)}{m r^{2m}} =  2\pi i \frac{\sqrt{-E/2}}{r} + \log r - \log(4\pi) - \frac12 \log(-E/2) + \log(1-e^{-\frac{2\sqrt2\pi i\sqrt{-E}}r}) - \frac{i \pi}{2}
\end{equation}
to get
\begin{align}
    \int_{E}^{\infty} &d\Delta\,\rho^{\text{scalars}}_{\text{MWK,}~E}(\Delta) f(\Delta, r) \nonumber \\&= \frac{1}{r - r^{-1}} \Bigg[\frac{2\pi i \sqrt{-E/2}}r + \log((1-e^{-2\sqrt2\pi i r\sqrt{-E}})(1-e^{-\frac{2\sqrt2\pi i \sqrt{-E}}r})) -\log(8\pi^2 \sqrt{-E \epsilon}) - \frac{i \pi}{2} \Bigg].
\end{align} 
This finally implies
\begin{align}
     \int_{E}^{\infty} d\Delta \,\rho^{\text{scalars}}_{\text{MWK,}~E}(\Delta) \left[f( \Delta,r) + f(\Delta, r^{-1})\right] &= 2\pi i \sqrt{-E/2}.
\end{align} 

Now we check MWK with a spinning seed. From (\ref{eq:keller76}), we have
\begin{align}
\int_E^\infty d\Delta \rho^{\text{scalars}}_{\text{MWK,}~E,~J}(\Delta&)\log\left(\frac{1-e^{-2\sqrt 2\pi r \sqrt{\Delta}}}{1-e^{-2\sqrt 2 \pi r^{-1} \sqrt{\Delta}}}\right) \nonumber \\
&= 4\sigma_0(J) \log r + 2\sum_{m=1}^\infty\frac{\sigma_{2m}(J) T_m(\frac{-E}J)}{J^m m} (r^{2m} - r^{-2m})
\end{align}
This sum of course diverges. However formally we have that
\be
\sum_{m=1}^\infty \frac{T_m(a)b^m}{m} = -\frac12 \log(b^2 - 2ab+1)
\label{eq:chebyshevtidentity}
\ee
Thus
\begin{align}
2\sum_{m=1}^\infty \frac{\sigma_{2m}(J) T_m\left(\frac{-E}J\right)}{J^m m} r^{2m} &= 2\sum_{m=1}^\infty \sum_{n | J} \frac{T_m\left(\frac{-E}J\right)}m \left(\frac{n^2 r^2}{J}\right)^m \nonumber \\
&= -\sum_{n|J} \log\left(\frac{n^4 r^4}{J^2} + \frac{2E n^2 r^2}{J^2} + 1\right).
\end{align}
Thus
\begin{align}
 2\sum_{m=1}^\infty\frac{\sigma_{2m}(J) T_m(\frac{-E}J)}{J^m m} (r^{2m} - r^{-2m}) &= \log \prod_{n|J} \frac{J^2 r^4 + n^4 + 2 r^2 n^2 E}{r^4(J^2 + r^2 n^2 (r^2 n^2+2E))} \nonumber \\
 &= -4 \sigma_0(J) \log r
\end{align}
(To show that
\be
 \prod_{n|J} \frac{J^2 r^4 + n^4 + 2 r^2 n^2 E}{r^4(J^2 + r^2 n^2 (r^2 n^2+2E))} = r^{-4\sigma_0(J)},
\ee
we pair up $n$ and $J/n$ in the product. Each pair gives a factor of $r^{-8}$.) 

Finally let us check the same but for a Rademacher sum as done in \cite{Alday:2019vdr}. We wrote down explicitly the scalar spectrum in (\ref{eq:aldaybaespec}).The first two terms come from the inverse Laplace transform of $e^{-2\pi E y} ~\text{erf}(\sqrt{-2\pi y E})$; the next two terms come from the $\frac{\log y}{\sqrt y}$ and $\frac{\text{constant}}{\sqrt y}$ pieces in $\frac{E_1}{\sqrt y}$; and the final term is all the $y^m$ pieces in the Eisensteins.

Now let us plug in (\ref{eq:aldaybaespec}) into our crossing equation. We can do each of the terms exactly. The first nontrivial term, for $E<0$, is
\begin{align}
    \int_0^\infty d\Delta &\left(-\frac{\sqrt{-E}}{\pi\sqrt\Delta (\Delta-E)}\right)\frac{\log\left(\frac{1-e^{-2\sqrt 2\pi r\sqrt{\Delta}}}{1-e^{-2\sqrt 2 \pi r^{-1} \sqrt{\Delta}}}\right)}{r - r^{-1}} = (2\sqrt 2-\sqrt 2 \log(-2E))\sqrt{-E} \nonumber \\&~~~~~~~-2\sqrt{-2E}\left(\frac{r+r^{-1}}{r-r^{-1}}\right) \log r - \frac{2\log r}{r-r^{-1} } - \frac{2\log\left(\frac{\Gamma(1+\sqrt{-2E}r^{-1})}{\Gamma(1+\sqrt{-2E} r)}\right)}{r-r^{-1}}.
\end{align}
We also have
\begin{align}
    \int_0^{\infty} d\Delta \log(1-e^{-2\sqrt 2 \pi r\sqrt{\Delta}}) \frac{\log\Delta}{\sqrt \Delta} = r^{-1}\left(\frac{\gamma_E \pi}{3\sqrt 2} - \frac{\zeta'(2)\sqrt 2}{\pi} + \frac{\pi \log(8\pi^2)}{6\sqrt 2} + \frac{\pi \log r}{3\sqrt 2}\right)
\end{align}
which implies
\begin{align}
  \int_0^\infty d\Delta \frac{12\sqrt{-E}}{\pi\sqrt\Delta}\log\Delta \frac{\log\left(\frac{1-e^{-2\sqrt 2\pi r\sqrt{\Delta}}}{1-e^{-2\sqrt 2 \pi r^{-1} \sqrt{\Delta}}}\right)}{r - r^{-1}} = 2\sqrt{-2E} \left(\frac{r^2+1}{r^2-1} \log r - \gamma_E - \log(2\sqrt 2\pi) + \frac{6 \zeta'(2)}{\pi^2}\right).
\end{align}
Finally the last term of (\ref{eq:aldaybaespec}) gives:
\begin{align}
    &4\pi\sum_{m=1}^\infty \frac{\zeta(2m+1)}{\zeta(2m+2)} \frac{16^m\pi^{2m}(-E)^{m+\frac12}}{\Gamma(2m+2)} 
 \int_0^\infty d\Delta \Delta^{m-\frac12} \frac{\log\left(\frac{1-e^{-2\sqrt 2\pi r\sqrt{\Delta}}}{1-e^{-2\sqrt 2 \pi r^{-1}\sqrt{\Delta }}}\right)}{r - r^{-1}} \nonumber \\
 &= \sum_{m=1}^\infty \frac{\zeta(2m+1)2^{m+\frac32}(-E)^{m+\frac12}}{2m+1} \left(\frac{r^{2m+1}-r^{-2m-1}}{r-r^{-1}}\right) \nonumber \\
 &= -2\sqrt{-2E}\gamma_E + \frac{\log\left(\frac{\Gamma(1+\sqrt{-2E} r^{-1}) \Gamma(1-\sqrt{-2E} r)}{\Gamma(1-\sqrt{-2E} r^{-1})\Gamma(1+\sqrt{-2E} r)}\right)}{r - r^{-1}} \nonumber \\
 &= -2\sqrt{-2E}\gamma_E + \frac{2\log r}{r-r^{-1}} + \frac{2\log\left(\frac{\Gamma(1+\sqrt{-2E}r^{-1})}{\Gamma(1+\sqrt{-2E} r)}\right)}{r-r^{-1}} - \frac{\log\left(\frac{\sin(\sqrt{-2E}\pi r)}{\sin(\sqrt{-2E}\pi r^{-1})}\right)}{r-r^{-1}}.
\end{align} 
Putting everything together, we get that
\begin{align}
    \int_{E}^\infty d\Delta \rho^{\text{scalars, Rad}}_E(\Delta)\left[f(\Delta, r) + f(\Delta, r^{-1})\right] &= 2\pi i \sqrt{-E/2} + \frac{\pi \mathfrak c\sqrt {-2E}}3 \nonumber \\ &~~~~~~+\sqrt{-2E}\left(\frac{12\zeta'(2)}{\pi^2}-3\gamma_E-\log(-2\pi E)+2\right) \nonumber \\
    &= 2\pi i \sqrt{-E/2},
\end{align}
where we used the definition of $\mathfrak c$ in (\ref{eq:mathfrakcdef}).

Note that \cite{Alday:2020qkm} also computed the Poincar\'{e} and Rademacher sums for a $\mathcal{W}_N$ algebra instead of a Virasoro algebra. This would be checked by slightly different scalar crossing equations, given in App. \ref{app:generalypower}. 

\section{Crossing equation with arbitrary power of $y$}
\label{app:generalypower}

Here we write the explicit generalizations of both (\ref{eq:maincrossing}) and (\ref{eq:crosswithzetastuff}) for a scalar partition function with an arbitrary power of $y^{w/2}$ multiplying the sum over states. Since the derivation uses identical logic as done in Sec. \ref{sec:freebosonintegral} and \ref{sec:Derivecross}, plus some identities derived in \cite{Benjamin:2022pnx, Benjamin:2023nts}, we will simply quote the results.

Consider a modular-invariant partition function and multiply by $y^{w/2}|\eta(\tau)|^{2w}$. We define $\rho^{\text{scalars}}_w(\Delta)$ as:
\begin{align}
    y^{w/2} \int_{-1/2}^{1/2} dx ~ |\eta(\tau)|^{2w} Z(\tau,\bar\tau) &\coloneqq y^{w/2} \int_{-\frac{c-w}{12}}^\infty d\Delta \rho^{\text{scalars}}_w(\Delta) e^{-2\pi y \Delta} \nonumber \\
    &\coloneqq Z^{\text{scalars}}_w(y)
\label{eq:zscalarssgendef}
\end{align}
Here $\rho^{\text{scalars}}_w(\Delta)$ counts the density of states as a function of the conformal weight, with $\frac{c-w}{12}$ subtracted, so for example the vacuum is at $\Delta=-\frac{c-w}{12}$.

We get a crossing equation of the form: 
\begin{equation}
    \int_{-\frac{c-w}{12}}^\infty d\Delta\, \rho^{\text{scalars}}_w(\Delta) (f_w(\Delta, r) + f_w(\Delta, r^{-1})) = \frac{\pi \varepsilon}6,
    \label{eq:gencrosssstuff}
\end{equation}
where 
\begin{equation}
    f_w(\Delta, r) \coloneqq \frac{2^{\frac{6-w}4}\Delta^{\frac{2-w}{4}}r^{\frac w2}}{r^{-1}-r}\sum_{m=1}^\infty m^{\frac{w-2}2} K_{\frac{w-2}2}\left(2\pi m \sqrt{2\Delta}r\right).
    \label{eq:fsgen}
\end{equation}
If we take (\ref{eq:fsgen}) at set $w=1$, it reduces to (\ref{eq:fs1def}); if we set $w=0$ it reduces to (\ref{eq:varepsnoystuff}) (which was also relevant in \cite{Benjamin:2023nts}). An integer value of $w$ is relevant for theories with $U(1)^w$ chiral algebra or $\mathcal{W}_{w+1}$ symmetry, but of course we can also set $w$ to be a non-integer value if we choose. Convergence of the integral (\ref{eq:gencrosssstuff}) requires $c<w+6(\text{min}(r,r^{-1}))^2 \leq w+6$.

We can also get a convergent crossing equation that also allows us to numerically extract $\varepsilon$ by using the techniques of Sec. \ref{sec:Derivecross}. We consider 
\begin{equation}
    \hat Z_w(\tau,\bar\tau) = Z_w(\tau,\bar\tau) -Z^{\text{gravity}}_w(\tau,\bar\tau),
\end{equation}
where we define $Z^{\text{gravity}}_{w}(\tau,\bar\tau)$ as
\begin{align}
    Z^{\text{gravity}}_{w}(\tau,\bar\tau) \coloneqq \sum_{\substack{E\leq 0,J}}\sum_{\gamma \in \Gamma_\infty \backslash SL(2,\mathbb Z)} \gamma(y^{w/2} q^{\frac{E+J}2}\bar{q}^{\frac{E-J}2}).
\end{align}
In particular, we will need the expressions for the scalar part $Z^{\text{gravity}}_{w;j=0}$,
\begin{equation}
    Z^{\text{gravity}}_{w;j=0}(y) = \int_{-1/2}^{1/2} dx\,\sum_{E\leq 0, J}\sum_{\gamma \in \Gamma_\infty \backslash SL(2,\mathbb Z)} \gamma(y^{w/2} q^{\frac{E+J}2}\bar{q}^{\frac{E-J}2}).
\end{equation}
For $J=0$, we have \cite{Maloney:2007ud,Keller:2014xba}:
\begin{align}
    \int_{-1/2}^{1/2} &dx \sum_{\gamma \in \Gamma_\infty \backslash SL(2,\mathbb Z)} \gamma(y^{w/2} q^{\frac{E}2}\bar{q}^{\frac{E}2}) = \int_{-1/2}^{1/2} dx \sum_{m=0}^{\infty} \frac{(-2\pi E)^m}{\Gamma(m+1)} E_{m+w/2}(\tau,\bar\tau) \nonumber\\
    &= y^{w/2} e^{-2\pi E y} + \sqrt\pi \sum_{m=0}^\infty \frac{(-2\pi E)^m \Gamma(m+\tfrac{w}{2}-\frac12) \zeta(2m+w-1)}{\Gamma(m+1)\Gamma(m+\tfrac{w}{2})\zeta(2m+w)} y^{1-m-\tfrac{w}{2}}.
\label{eq:j0sumarbitrary}
\end{align}
In (\ref{eq:j0sumarbitrary}), if ever there is a term such that $m+\tfrac{w}{2}=1$, we take the limit as $m$ approaches $1-\tfrac{w}{2}$ and keep the non-singular term, just as in (\ref{eq:limitnonsingular}). For example, if $w=0$, then the $m=1$ term in (\ref{eq:j0sumarbitrary}) is singular; by taking the limit, we recover (\ref{eq:statesdensityspinJ0}).

For $J\neq0$, we have \cite{Maloney:2007ud, Keller:2014xba}:
\begin{align}\label{eq:spinningsumarbitrarypower}
    \int_{-1/2}^{1/2} dx \sum_{\gamma \in \Gamma_\infty \backslash SL(2,\mathbb Z)} \gamma(y^{w/2} q^{\frac{E+J}2}\bar{q}^{\frac{E-J}2}) &= \sum_{m=0}^{\infty} \frac{(-2\pi E)^m}{\Gamma(m+1)} \int_{-1/2}^{1/2} dx\, E(m+\tfrac{w}{2},\tau,\bar\tau,J),
\end{align}
where $E(m+\tfrac{w}{2},\tau,\bar\tau,J)$ is given explicitly in (E.5) of \cite{Benjamin:2021ygh}.

The spectral decomposition for $Z^{\text{scalar}}_w$ is given by\footnote{Potentially, $Z^{\text{gravity}}_{w}$ could give rise to polynomially growing terms that are faster than $y^{1/2}$ at large $y$, which will also need to be subtracted out. For $w\in \mathbb{Z}_{\geq 0}$, we do not have such terms.}
\begin{align}\label{eq:Vircross_generalw_expr0}
&Z^{\text{scalars}}_w(y) - Z^{\text{gravity}}_{w;j=0}(y)=\frac{3}{\pi} (\hat Z_w,1)+\frac{1}{2\pi i}\int_{\mathrm{Re}(s)=\frac{1}{2}}ds\, (\hat Z_w,E_s)y^s.
\end{align}
After deforming the contour, the final equation we get is:
\begin{align}
    &Z^{\text{scalars}}_w(y) - Z^{\text{gravity}}_{w;j=0}(y) \nn \\
    &= \frac{3}{\pi} (\hat Z_w,1) + \sum_{k=1}^\infty \text{Re}\left(\delta_k y^{\frac{1+z_k}2}\right) \nonumber \\
    &+ \frac{y^{1-\frac w2}}{\sqrt\pi} \sum_{n=1}^\infty b(n) n^{w-2} \int d\De\, (\rho^{\text{scalars}}_{0}(\De) - \rho^{\text{gravity}}_{w;j=0}(\De)) U\left(-\frac12, \frac w2, \frac{2\pi n^2 \Delta}y\right) e^{-\frac{2\pi n^2\Delta}y}.
    \label{eq:genzetaarbs}
\end{align}
In (\ref{eq:genzetaarbs}), $Z^{\text{scalars}}_w(y)$ is defined in (\ref{eq:zscalarssgendef}), $z_k$ are the nontrivial zeros of the Riemann zeta function, $b(n)$ is defined in (\ref{eq:bndef}), $U$ is a confluent hypergeometric function of the second kind, and $\rho^{\text{scalars}}_{0}(\De)$ is the density of states of all scalar operators. (Noet that after subtracting $\rho^{\text{gravity}}_{w;j=0}(\De)$, only operators with $\De>0$ will contribute, which means the last line of (\ref{eq:genzetaarbs}) is nonperturbatively small in $y^{-1}$.) $Z^{\text{gravity}}_{w;j=0}$ is defined as the scalar part of the Poincar\'{e} sum of all operators with $\Delta \leq 0$, with an overall power of $y^w$. More explicitly, there are a finite number of operators of energy $E\leq 0$ and spin $J$. Finally, $\rho^{\text{gravity}}_{j=0}(\De)$ is defined by the equation
\begin{align}
    Z^{\text{gravity}}_{w;j=0}(y) = y^{\frac{w}{2}}\int d\De\, \rho^{\text{gravity}}_{w;j=0}(\De) e^{-2\pi \De y}.
\end{align}

The contribution from $\rho^{\text{gravity}}_{j=0}(\De)$ can be evaluated explicitly. Setting $w=1$ of (\ref{eq:genzetaarbs}) then reduces to (\ref{eq:crosswithzetastuff}), and setting $w=0$ allowed us to do the calculations in Sec. \ref{sec:modularzetas0}. The crossing equation for $w=0$ is given by
\begin{align}\label{eq:crosiingwithzeta_states}
     &Z^{\text{scalars}}_{w=0}(y) - Z^{\text{gravity}}_{w=0;j=0}(y) \nonumber \\
    &= \frac{3}{\pi} (\hat Z_0,1) + \sum_{k=1}^\infty \text{Re}\left(\delta_k y^{\frac{1+z_k}2}\right) \nonumber \\
    &+\sum_{\Delta\leq 0,J=0} (1-e^{-2\pi \Delta y}) \nonumber \\
    &+\sum_{\Delta\leq 0,J\neq 0}  \sum_{a,s=0}^{\infty}\!\!{}^\prime \frac{(-1)^{a+s}(-2\pi \Delta)^{s}\pi^{2a-1}\Gamma(\tfrac{3}{2}-2a-s)\Gamma(-\tfrac{1}{2}+a+s)\sigma_{4a+2s-1}(|J|)}{|J|^{2a+2s-1}\Gamma(a+1)\Gamma(s+1)\zeta(4a+2s-1)}y^{2a+s} \nonumber \\
    &+ \frac{y}{\sqrt\pi}\sum_{\Delta > 0, J=0} \sum_{n=1}^\infty b(n) n^{-2}  U\left(-\frac12, 0, \frac{2\pi n^2 \Delta}y\right) e^{-\frac{2\pi n^2\Delta}y},
\end{align}
where for the sum $\sum_{a,s=0}^{\infty}\!\!{}^\prime$ we do not include the $a=s=0$ term or the $a=0,s=1$ term.

Finally, we can apply the functionals introduced in section \ref{sec:functional_sumrules} to \eqref{eq:crosiingwithzeta_states}. This gives the equation \eqref{eq:varepsnoystuff}, where $\varepsilon$ is now given by
\begin{equation}
    \varepsilon=\frac{3}{\pi}(\hat Z_0,1)+\sum_{\Delta\leq 0,J=0}(1-6\pi \Delta i)-\sum_{\Delta\leq 0,J\neq 0}12 \sigma_1(|J|).
\end{equation}
The last two terms can be viewed as the $\varepsilon$ for MWK partition function with scalar and spinning seeds (see \eqref{eq:epsilonsmwkw0}).

\bibliographystyle{JHEP}
\bibliography{ScalarVir}
\end{document}